\documentclass[11pt]{article}
\usepackage{amsmath}
\usepackage{amssymb}
\usepackage{amsthm}
\usepackage{graphicx}
\usepackage{natbib}
\usepackage{booktabs}
\usepackage[left=25.4mm,right=25.4mm,top=30mm,bottom=25.4mm,a4paper]{geometry}
\usepackage{caption}
\usepackage{subcaption}

\theoremstyle{definition}

\newcommand{\E}{\mathbb E}
\newcommand{\e}{\mathrm e}

\newcommand{\D}{\mathrm{d}}

\newcommand{\I}{\mathrm i}

\begin{document}

\title{Performance of tail hedged portfolio with third moment variation swap}
\author{Kyungsub Lee\footnote{Department of Statistics, Yeungnam University, Gyeongsan, Gyeongbuk 38541, Korea}\thanks{This work was supported by the 2015 Yeungnam University Research Grant.}
and Byoung Ki Seo\footnote{School of Business Administration, UNIST, Ulsan 44919, Korea}
\thanks{Byoung Ki Seo was supported by the 2012 Research Fund(1.120071.01) of UNIST(Ulsan National Institute of Science and Technology).}
}

\date{}

\maketitle

\begin{abstract}
The third moment variation of a financial asset return process is defined by the quadratic covariation between the return and square return processes.
The skew and fat tail risk of an underlying asset can be hedged using a third moment variation swap under which a predetermined fixed leg and the floating leg of the realized third moment variation are exchanged.
The probability density function of the hedged portfolio with the third moment variation swap was examined using a partial differential equation approach.
An alternating direction implicit method was used for numerical analysis of the partial differential equation.
Under the stochastic volatility and jump diffusion stochastic volatility models, the distributions of the hedged portfolio return are symmetric and have more Gaussian-like thin-tails.
\end{abstract}

\section{Introduction}

The distribution of a financial asset return is negatively skewed and has fat tails compared to a normal distribution.
For risk management, asset pricing and hedging purposes, it is important to consider the high moments of the return distribution.
Despite their importance, the third and fourth moments of asset returns are difficult to measure precisely by averaging the third and fourth powers of the sample returns due to the large deviations in the estimators.

One of the methods for estimating the third moment of the return distribution is to use the third moment variation process of the return based on high-frequency data.
The third moment variation is defined as the quadratic covariation between the return and its squared processes over a fixed time period \citep{ChoeLee}.
This approach is an extension of the growing literature regarding the realized variance of return, including \cite{Barndorff2002a},\cite{ABDL}, \cite{Barndorff-Nielsen2004}, \cite{Hansen}, \cite{MyklandZhang} and \cite{Wang}.
Similarly, with the realized variance, the third moment variation has good properties as an estimator such as consistency, relative efficiency and unbiasedness under a martingale assumption \citep{Lee2015}.

Furthermore, the third moment variation can play an important role in hedging the skew and fat tail risks of return distributions.
An investor who wants to hedge the skew and tail risk may contract the third moment variation swap under which a predetermined fixed leg and the negative value of the realized third moment variation, the floating leg, are exchanged at maturity.
The basic trading mechanism of the third moment variation swap is similar to the variance swap.
If the underlying asset price plunges, then the floating leg is likely to have a positive value,
since the third moment variation itself is likely to have a negative value in the turmoil, 
and the floating leg of the swap is defined as the negative value of the third moment variation.
Therefore, an investor can be compensated for the loss of an underlying asset by the floating leg of the swap.
As a result, the return of the total portfolio follows a more Gaussian-like thin-tail distribution than the underlying asset alone.

Similar studies of trading the skew risk have been reported.
\cite{Schoutens} defined the moment swaps based on the finite sum of $k$-th powers of log-return and showed that the hedging performance of the variance swap can be enhanced by a third moment swap, where the third moment swap is defined as being different from our approach.
\cite{Neuberger2012} constructed a similar approach for a skew swap but this was also based on different definition of the measure of skewness focused on the aggregation property.
With this definition of the skew swap, \cite{Kozhan} examined the skew risk premium in equity index markets.
For more information on financial studies about skewness in asset returns, see \cite{KrausLitzenberger}, \cite{HarveySiddique1999}, \cite{HarveySiddique}, \cite{BakshiKapadiaMadan} and \cite{Christoffersen}.

This study examine the return distribution of the portfolio hedged by the third moment variation swap based on partial differential equations.
With the definition of the third moment variation swap and under the stochastic volatility and jump diffusion model, the joint probability density function of the hedged portfolio and underlying return is represented by partial differential equations.
The probability density function is calculated using the alternating direction implicit methods (ADI).
The ADI scheme is an efficient algorithm for a numerical solution to partial differential equations and there are financial applications such as~\cite{Hout}, \cite{Haentjens}, \cite{Jeong} and \cite{Haentjens2015}.
The numerical results show that the third moment variation swap hedges the fat tails of the underlying asset return distributions in terms of skewness and kurtosis under both the stochastic volatility and stochastic volatility with jump models.

The remainder of the paper is organized as follows:
Section~\ref{Sect:Empirical} explains the structure of the third moment swap and examine the empirical performance based on the S\&P 500 return series.
In Section~\ref{Sect:PDE}, the probability density function is computed under the stochastic volatility model.
Section~\ref{Sect:JD} extends the result to a jump diffusion stochastic volatility model.
Section~\ref{Sect:concl} concludes the paper.

\section{Third moment variation swap}\label{Sect:Empirical}

\subsection{Basic mechanism}
This subsection briefly reviews the structure of the third moment swap to hedge the skew and tail risk, as introduced by \cite{ChoeLee}, and examines the hedging performance using the S\&P 500 index.
The swap is based on the quantity called the third moment variation of the return process.
(More precisely, it is a covariation but for simplicity, it is called the third moment variation.)
The third moment variation of a semimartingale return process $R$ is defined by the quadratic covariation between the return and its square processes as follows:
\begin{align*}
[R,R^2]_t 
=  \lim_{\|\pi_n \| \rightarrow 0} \sum_{i=1}^{N} (R_{t_i} - R_{t_{i-1}})(R^2_{t_i} - R^2_{t_{i-1}}) \quad \textrm{in probability} 
\end{align*}
where $\pi_n$ is a sequence of partitions $0 = t_0< \cdots < t_N =t$ and $\|\pi_n\|$ is the mesh of the partition.

In addition,
\begin{align*}
[R,R^2]_t &= [R,R^2]^c_t + \sum_{0<s\leq t} \Delta R_s \Delta \left( R^2_s \right)\\
&= 2\int_0^t R_{u-} \D [R]^c_{u} + \sum_{0<s\leq t} \Delta R_s \Delta \left( R^2_s \right)
\end{align*}
where the superscript $c$ denotes the continuous part of the corresponding process.
For the second equality, the following is used: 
$$ R^2_t = 2 \int_0^t R_{u-} \D R_u + [R]_t$$
and
the fact that the continuous part of $[R,R^2]$ is the continuous part of the covariation between $2 \int_0^t R_{u-} \D R_u$ and $R_t$, i.e., $2\int_0^t R_{u-} \D [R]^c_{u}$.
This mathematical definition associated with the stochastic integration is in line with \cite{protter2013}.
The fourth moment variation of the return is similarly defined as $[R^2]$ but this paper focuses on the third moment variation.

The payoff structure of the third moment variation swap contract is similar to the variance swap where the predetermined fixed leg and realized variance over a certain time period is exchanged at maturity.
The different part from the variance swap is that the floating leg of the third moment swap is the negative value of the realized third moment variation, $-[R,R^2]_T$, over the period $[0,T]$.
As in the variance swap, the buyer of the swap pays the fixed leg and receives the floating leg.

Note that the third moment variation can have positive or negative values.
In a market plunge, the third moment variation of the underlying return are	likely to have a negative value;  hence, the floating leg of the swap, which is the minus of the variation, is likely to have a positive value.
In contrast, if a stock price rises sharply, then the third moment variation tends to have a positive value and hence the floating leg of the swap tends to have negative value.
Therefore, when an investor experiences a huge loss from an underlying asset price plunge, they can be compensated for the loss by the swap.
When an investor earns a huge profit from the long position of an underlying asset, they pays the floating leg to the counterparty of the swap contract as a kind of insurance.
Although the third moment variation swap has a simple mechanism, one can hedge the skew and fat-tail risk of the underlying asset by contracting the swap.

\begin{figure}
        \centering
        \begin{subfigure}[b]{0.4\textwidth}
                \includegraphics[width=\textwidth]{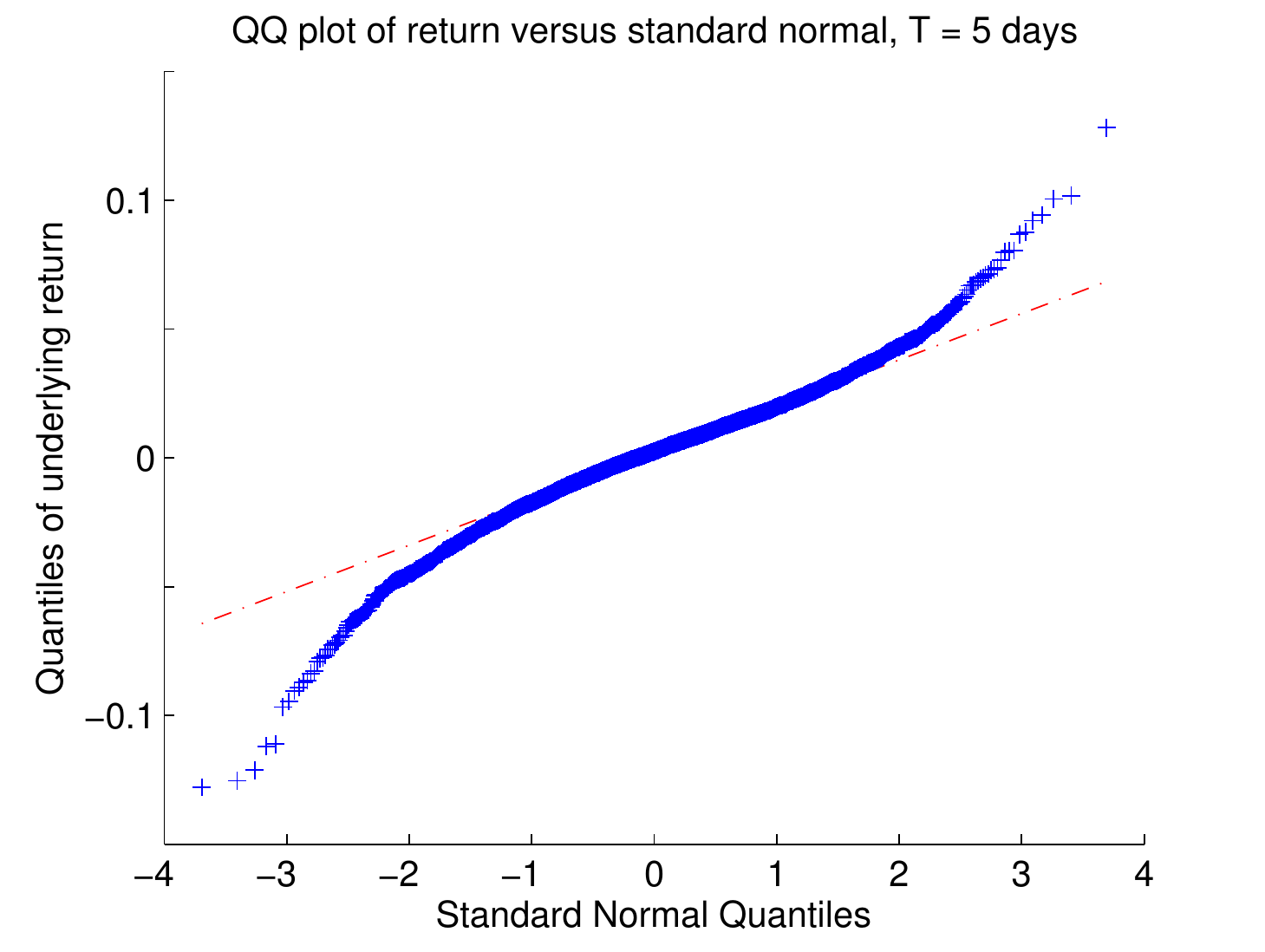}
                \caption{}
                \label{fig:QQS5}
        \end{subfigure}
  		\quad
        \begin{subfigure}[b]{0.4\textwidth}
                \includegraphics[width=\textwidth]{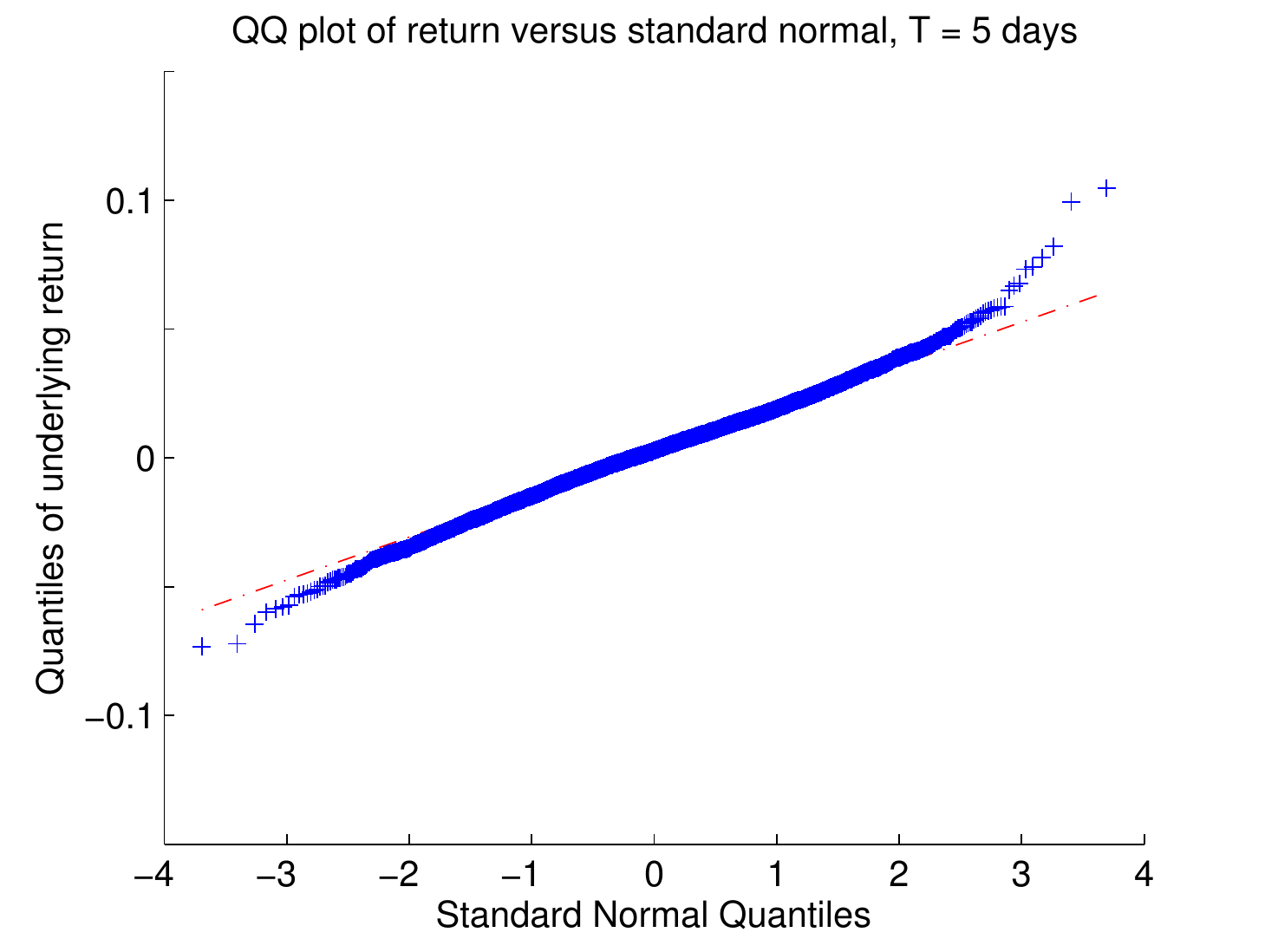}
                \caption{}
                \label{fig:QQP5}
        \end{subfigure}
        
        \begin{subfigure}[b]{0.4\textwidth}
                \includegraphics[width=\textwidth]{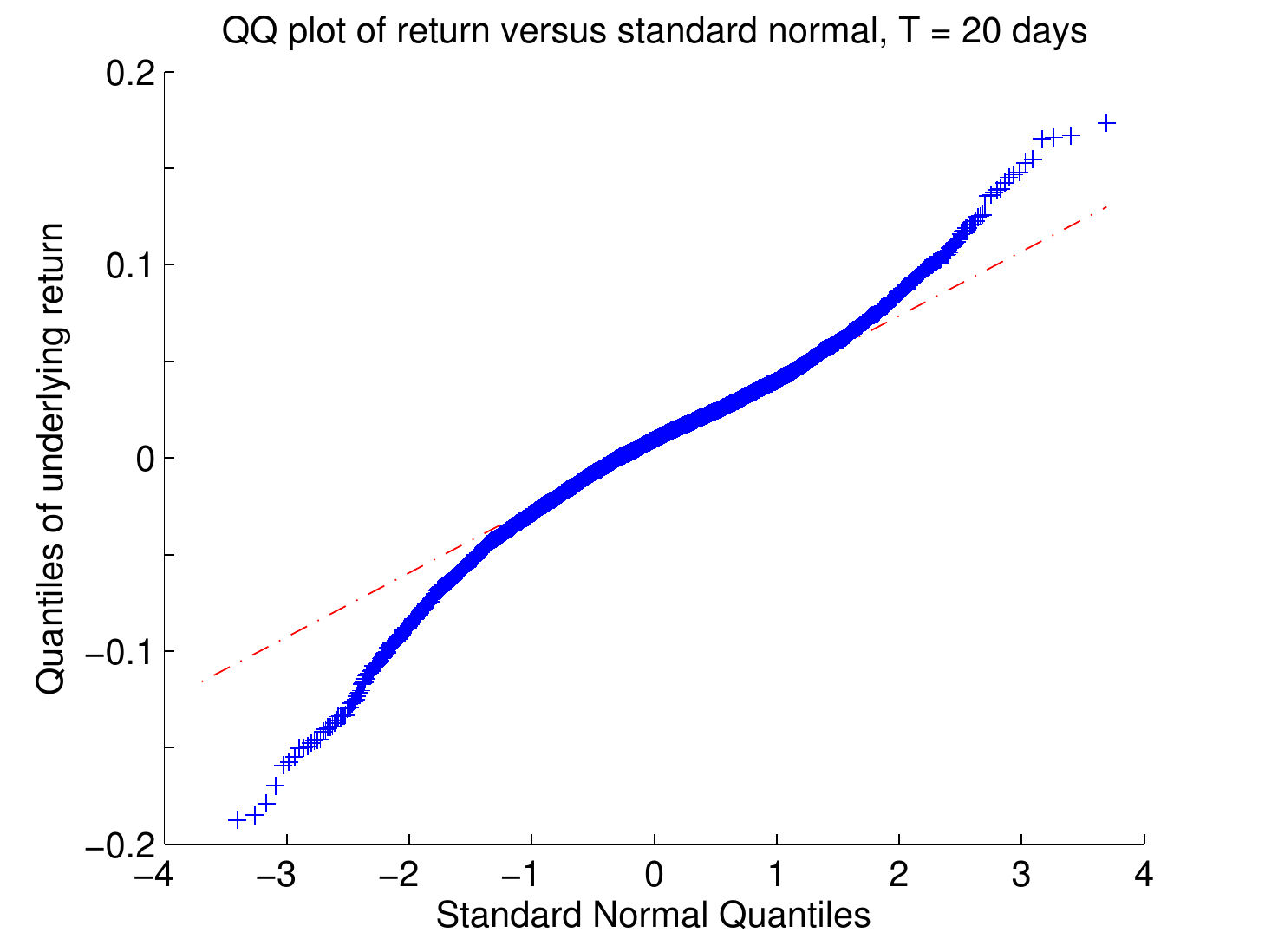}
                \caption{}
                \label{fig:QQS20}
        \end{subfigure}
  		\quad
		\begin{subfigure}[b]{0.4\textwidth}
                \includegraphics[width=\textwidth]{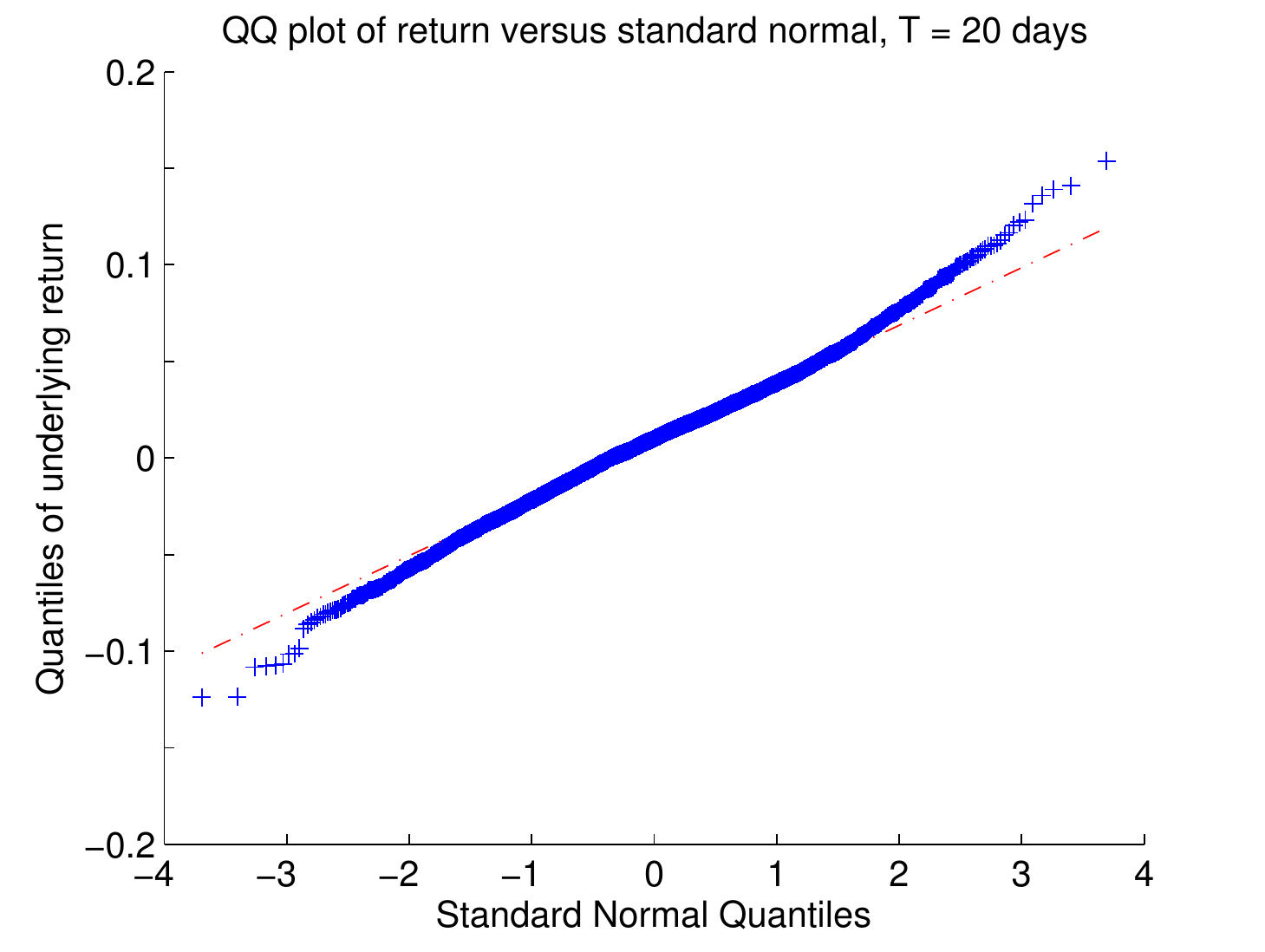}
                \caption{}
                \label{fig:QQP20}
        \end{subfigure}
        
		\begin{subfigure}[b]{0.4\textwidth}
                \includegraphics[width=\textwidth]{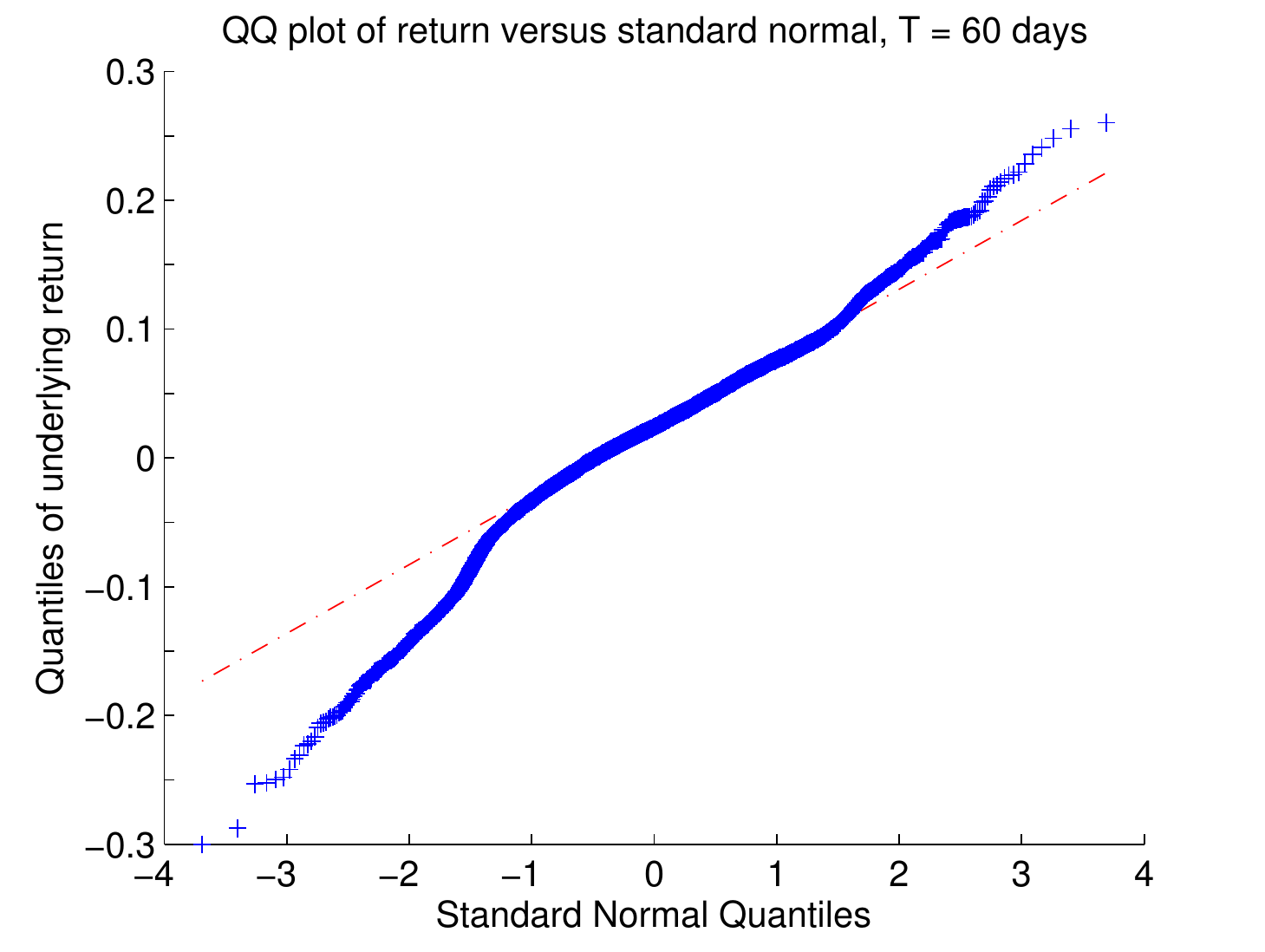}
                \caption{}
                \label{fig:QQS60}
        \end{subfigure}
  		\quad
		\begin{subfigure}[b]{0.4\textwidth}
                \includegraphics[width=\textwidth]{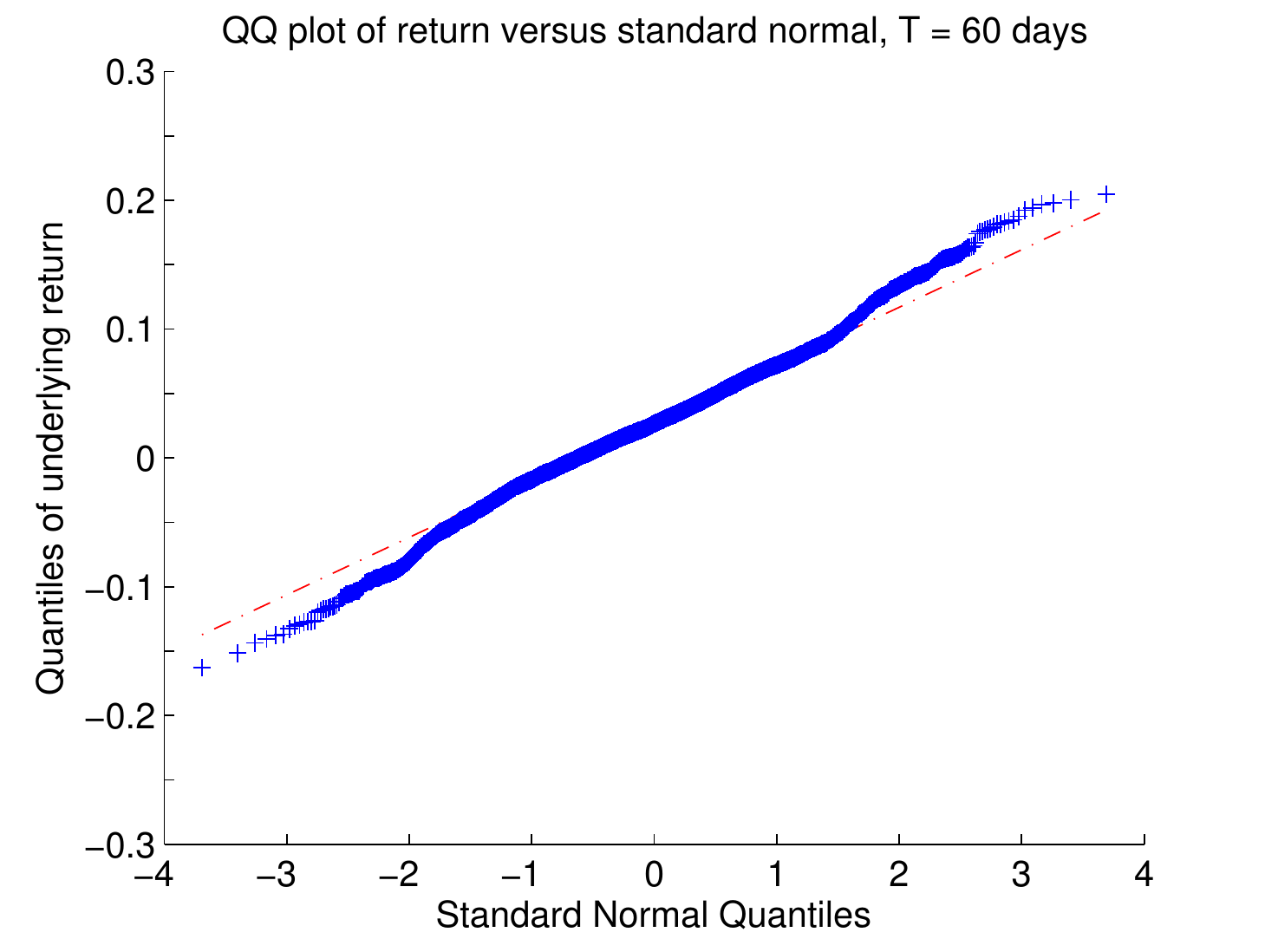}
                \caption{}
                \label{fig:QQP60}
        \end{subfigure}
        
		\begin{subfigure}[b]{0.4\textwidth}
                \includegraphics[width=\textwidth]{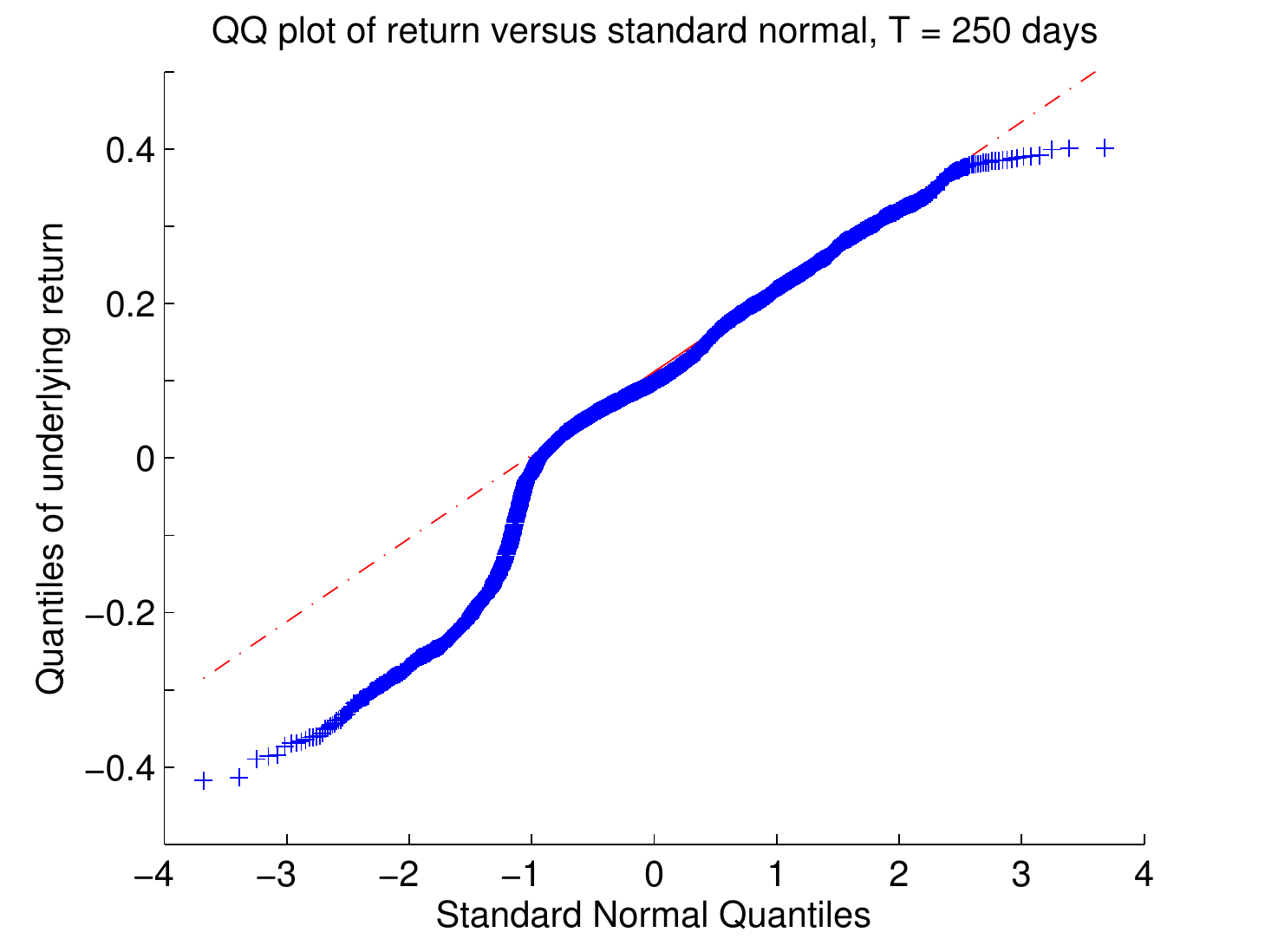}
                \caption{}
                \label{fig:QQS250}
        \end{subfigure}
  		\quad
		\begin{subfigure}[b]{0.4\textwidth}
                \includegraphics[width=\textwidth]{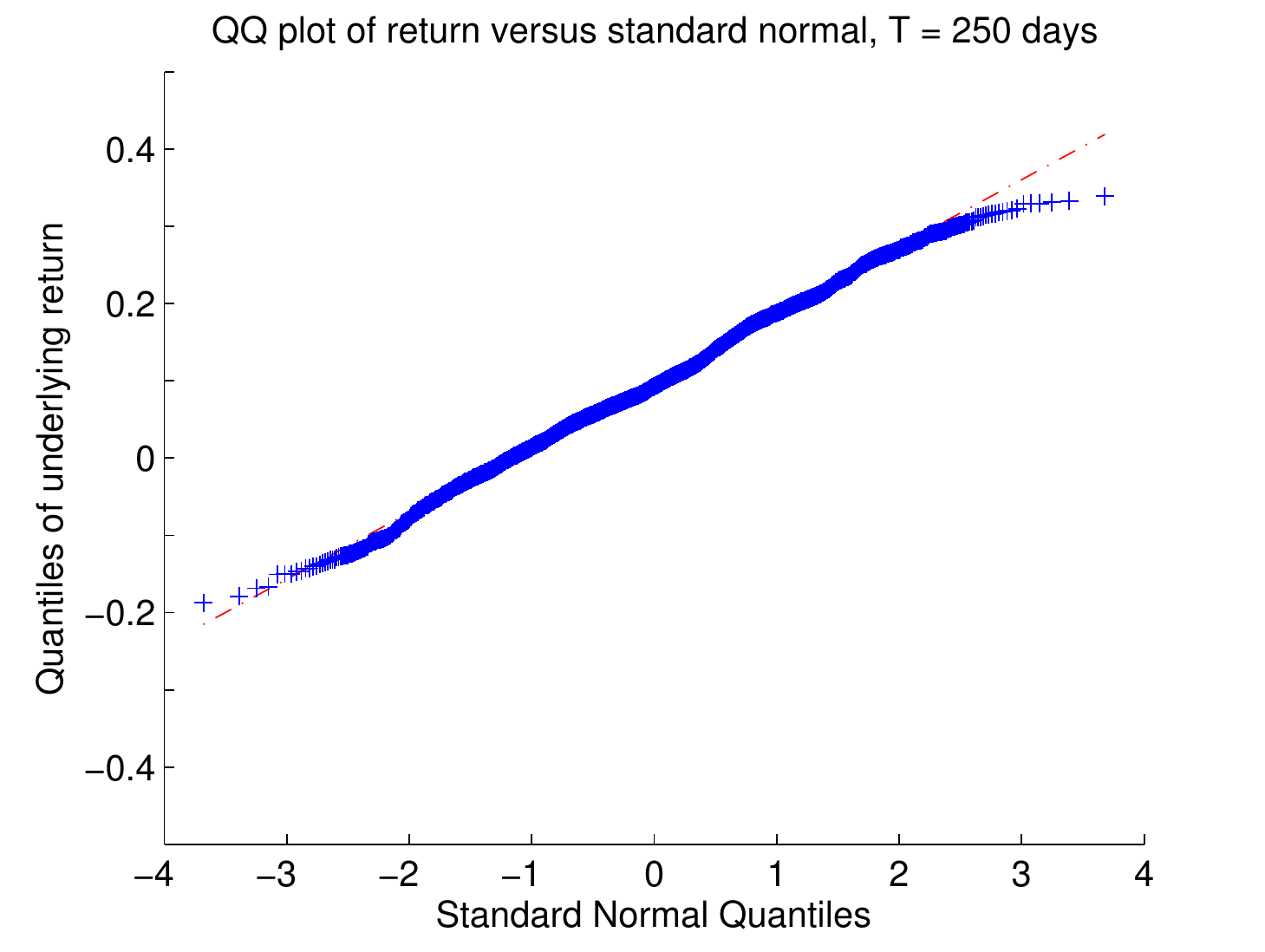}
                \caption{}
                \label{fig:QQP250}
        \end{subfigure}
        \caption{QQ plots of the underlying asset returns (left) and the hedged portfolios' returns (right) with various maturities}\label{fig:QQ}
\end{figure}

With a properly chosen number of swap amounts, the probability distribution of the portfolio consisting of the underlying asset and the swap exhibits a more Gaussian-like thin tail distribution.
An empirical study was employed using the S\&P 500 index five-minute data ranging from 1990 to 2007.
The left panels in Figure~\ref{fig:QQ} illustrate the quantile-quantile (QQ) plots of the S\&P 500 index return with various time lengths, $T = 5, 20, 60$ and $250$ work-days. 
All plots show a significant negative fat tail compared to the normal distribution.
The right panel shows the QQ-plots of the hedged portfolio by the third moment swaps with previously mentioned maturities, and the plots represent thinner return distributions.

In this analysis, for simplicity, the fixed values of the swaps are assumed to be zeros.
Indeed, the fixed value of the swap is not necessarily zero and its fair price would be determined by the market participants.
The theoretical values of the swap are expected to be based on European option prices 
under the assumption that the option prices properly reflect the risk-neutral measure.
For more information on the pricing issues of the swap, see~\cite{ChoeLee}.
Note that the fixed value of the swap, i.e., the price of the swap, is predetermined at the moment of contract 
and only affects the mean of the portfolio's return distribution, but not other distributional properties such as the variance, skewness or kurtosis.
Therefore, without a loss of generality, the fixed value of the swap is set to zero 
because this study focus on the shape of the return distribution of the hedged portfolio.

The hedge numbers are determined to minimize the squares of the differences between the quantiles of the realized portfolio return, hypothetically hedged by the swap, and the normal distribution function with the same mean and standard deviation with the portfolio distribution.
In other words, the $L_2$-norm of the difference between the empirical portfolio return and the corresponding normal distribution was minimized.
The amounts of the swap positions are $242.9, 80.8, 46.2$ and $16.2$ multiplied by the initial index values for various maturities $T = 5, 20, 60$ and $250$ days, respectively.
For example, if the index is 1,000 at time zero and the maturity of the swap is 20 work days,
then the notional amounts of the swap position for a unit index is $80.8 \times 1,000 = 80,800$.
This shows that the third moment variation swaps hedge the skew and fat tail risk so that the portfolio returns tend to have a more Gaussian-like distribution for all time intervals.

\begin{figure}
\centering
\includegraphics[width=0.45\textwidth]{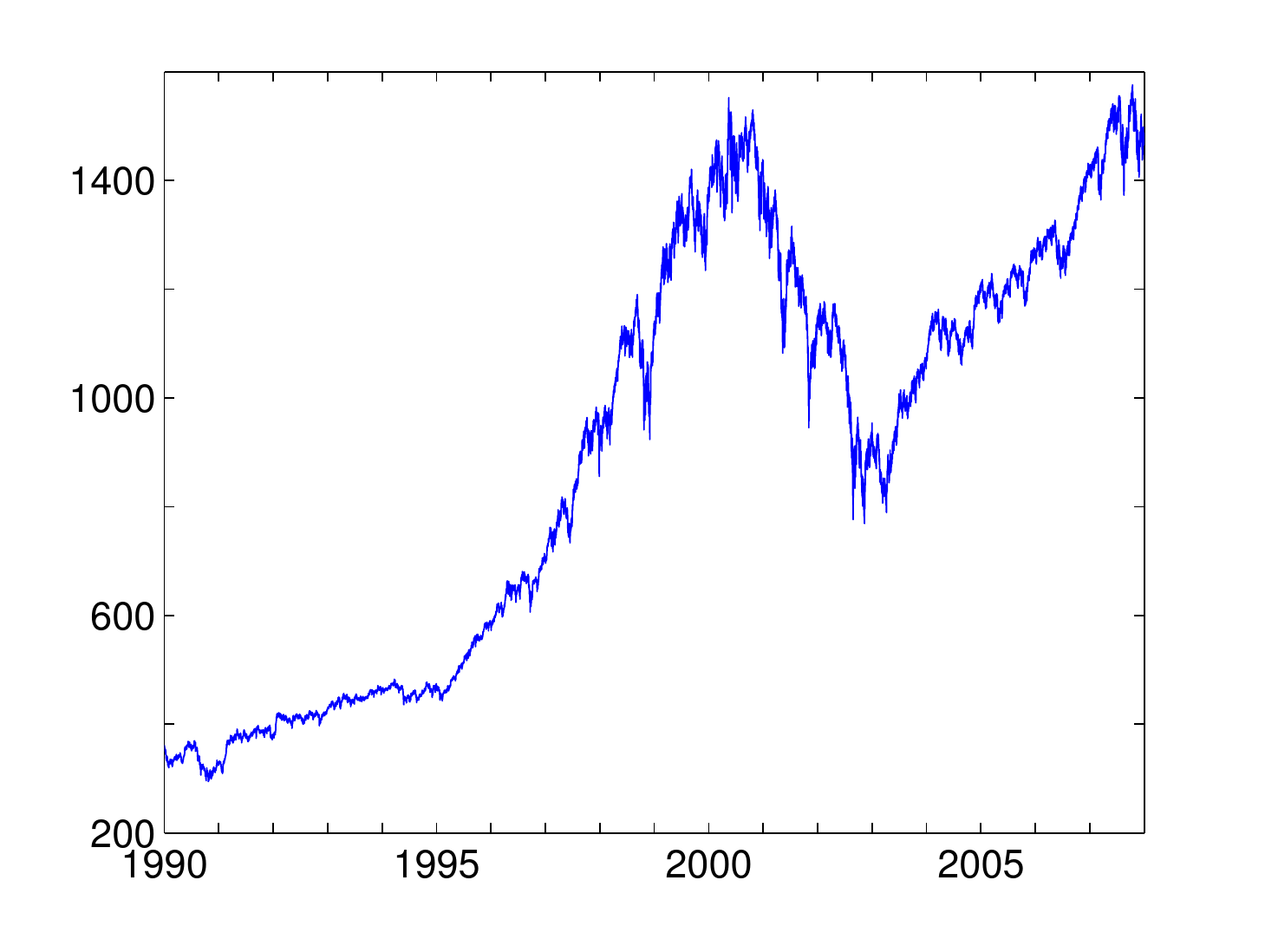}
\quad
\includegraphics[width=0.45\textwidth]{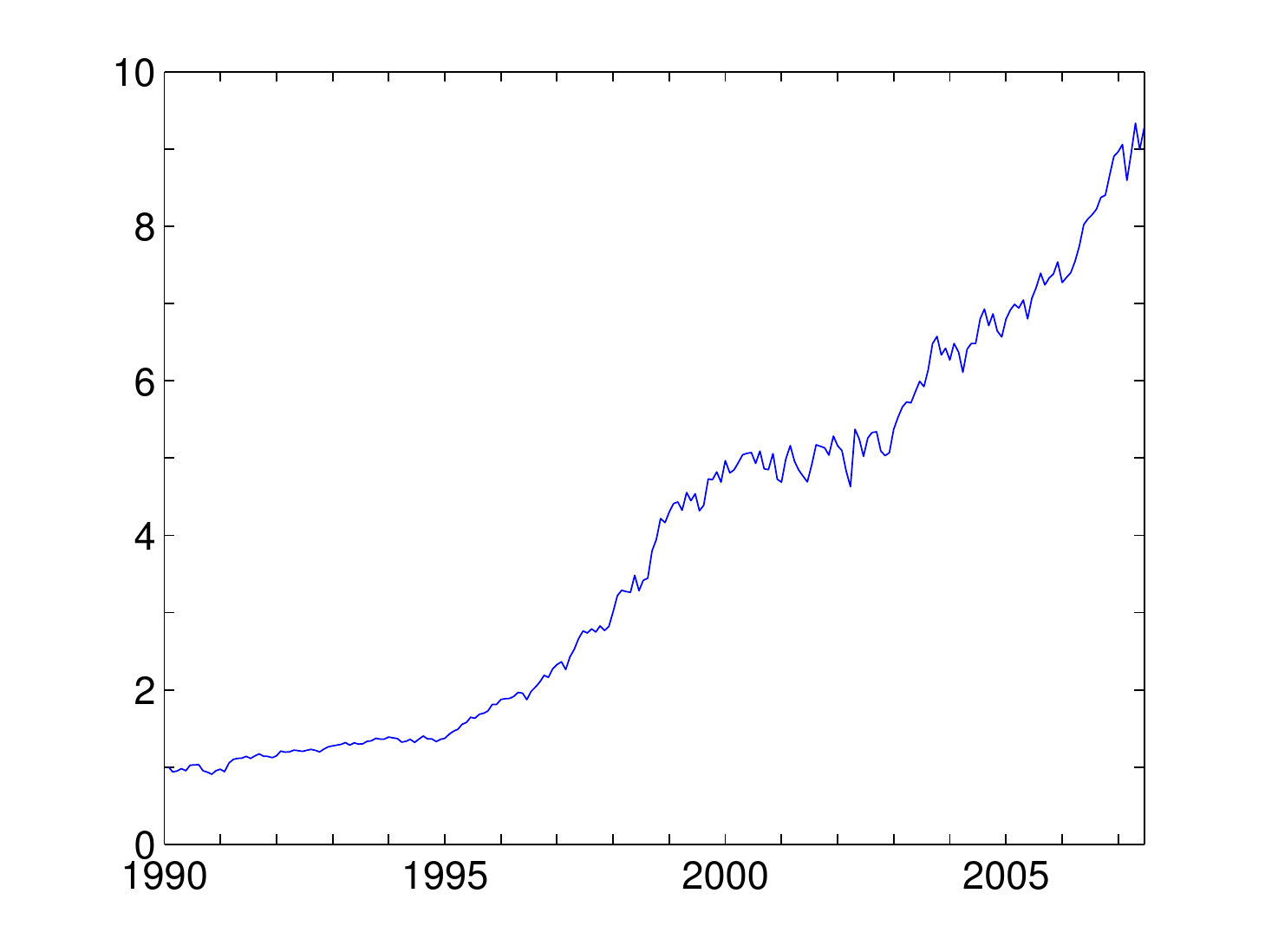}
\caption{The dynamics of S\&P 500 index (left) and the hedged portfolio' value (right) from 1990 to 2007}\label{Fig:dynamics}
\end{figure}

In the left of Figure~\ref{Fig:dynamics}, the dynamics of the S\&P 500 index from 1990 to 2007 is plotted.
The cumulative profit and loss of the hedged portfolio with the third moment variation swap where the investor rebalance the portfolio every month, i.e., the investor contracts a one-month third moment variation swap at January 1st, 1990, and contracts a new swap at the first day of the subsequent month etc., is presented in the right of the figure.
Compared to the index dynamics, the return of the hedged portfolio shows more steady growth all over time, even during the dot-com bubble crash in the early 2000's.
The relatively large excess return in the hedged portfolio compared to the S\&P 500 index is due to the assumption that the fixed leg of the swap is zero.

\subsection{Transaction cost, wrong-way and counterparty risk}

\begin{figure}
\centering
\includegraphics[width=0.45\textwidth]{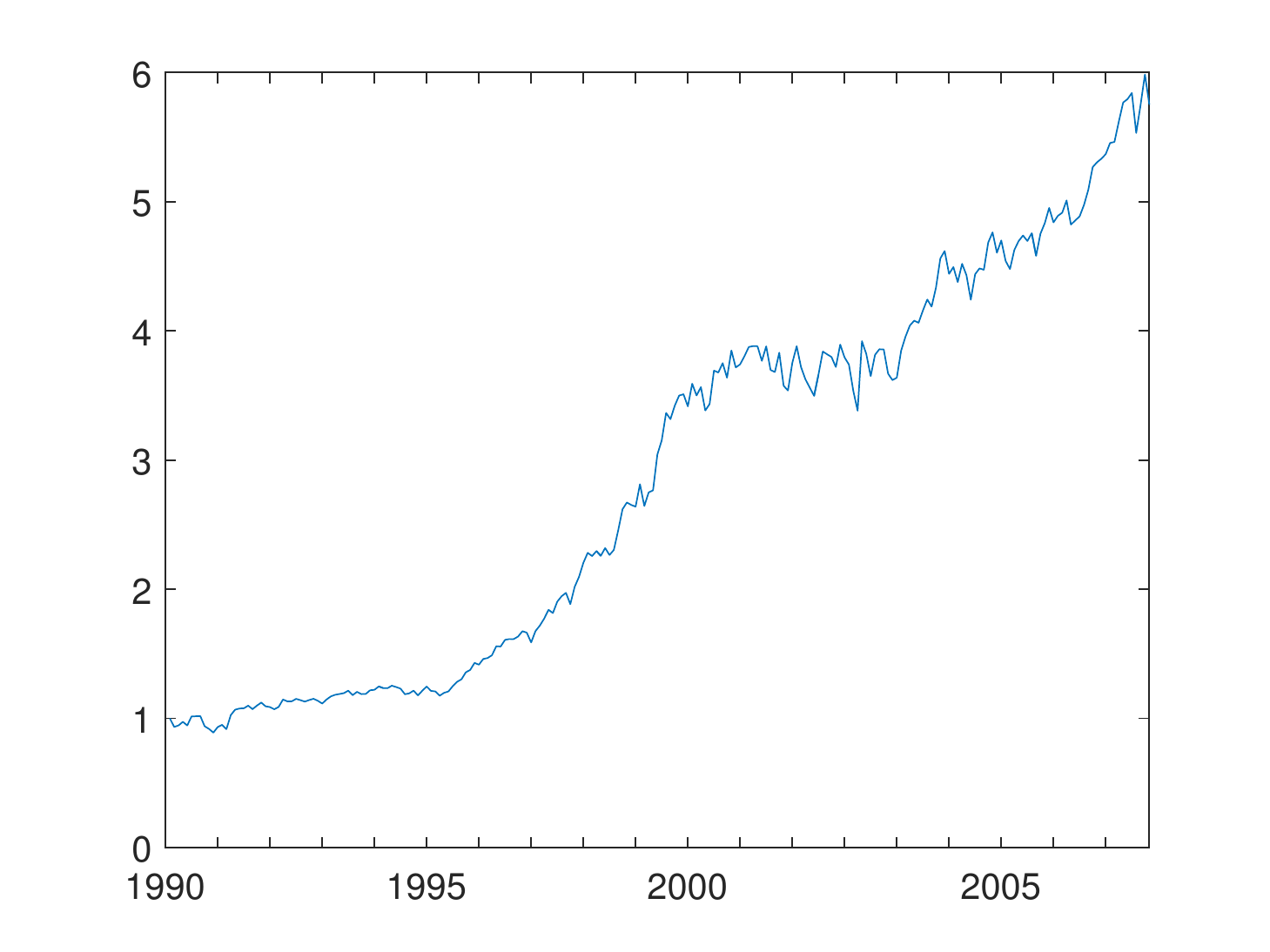}
\quad
\includegraphics[width=0.45\textwidth]{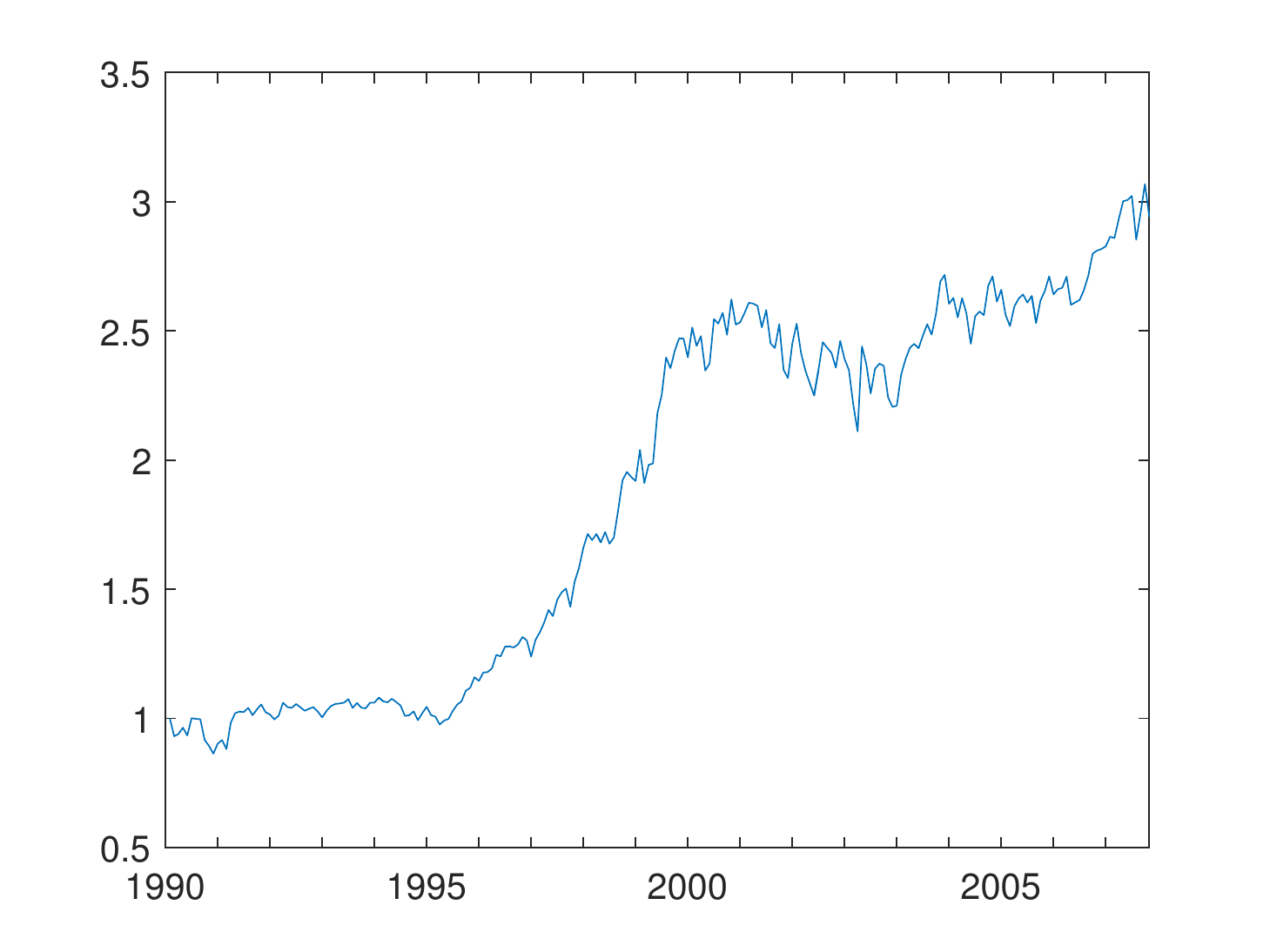}
\caption{The dynamics of the hedged portfolios' values with transaction cost 0.2\% (left) and 0.5\% (right)}
\label{Fig:cost}
\end{figure}

Consider the effect of the transaction cost such as bid-ask spread due to the illiquidity of the third moment variation swap.
The transaction cost does not affect the shape of the hedged portfolio's conditional distributions upon the contract date,
since the transaction cost is predetermined at the time of the contract.
The total return of the hedged portfolio over the long run period, e.g., three years, will be diminished,
when one repeatedly contract the third moment variation swap, e.g., every month as in the previous example.
The dynamics of the hedged portfolios' values are plotted along with presumed transaction costs with 0.2\% and 0.5\% of the underlying asset price in Figure~\ref{Fig:cost} based on the S\&P 500 index.
As expected, with large amounts of transaction costs, the total returns have been diminished from 1990 to 2007.

One way to avoids the risk associated with the transaction costs is to contract a long-term third moment swap with multiple legs similar to the interest rate swap.
For example, one can contract the third moment variation swap with three years maturity and the legs with a one month interval.
The swap then has 36 floating and fixed legs to be exchanged and the exchange of each leg is performed in the same way as the single third moment variation swap explained in the previous subsection.
In this way, the buyer of the swap can be compensated for the loss by the skew distribution every month for up to three years without taking the risks associated with future possible transaction costs.

The third moment variation swap can transfer the skew and tail risk from one party to another but cannot remove the risk entirely.
In addition, in market turmoil, the swap seller who has an obligation to pay the floating leg to the buyer might have difficulty in making payment.
The swap is associated with a wrong-way risk as the buyer's exposure to the counterparty is correlated with the seller's credit risk.
If the underlying asset price plunges, then the swap seller's payment to the buyer tends to increase 
but in such an asset price crash, the seller is also likely to be exposed to severe market and credit risk.
One way to minimize the wrong-way and counterparty risk is central clearing.
Central counterparties (CCP) bear the counterparty credit risk of the bilateral trades, such as the interest rate and credit default swaps, 
and the role of CCP becomes increasing worldwide.
The third moment variation swap also can be standardized and hence expected to be traded via CCPs.

\section{Partial differential equations}\label{Sect:PDE}

In this section, the probability density function of the tail hedged portfolio with the third moment variation swap under a stochastic volatility model is computed.
Since the analytical formula of the distribution of the tail hedged portfolio is not known,
this paper proposes a PDE and a numerical approach to calculate the probability density function of the tail hedged portfolio.
Assume that the asset return process $R$ follows the square root stochastic volatility model (as reported by \cite{Heston1993}):
\begin{align*}
\D R_t &= \left( \mu - \frac{1}{2}V_t \right) \D t + \sqrt{V_t} \D W^s_t \\
\D V_t &= \kappa (\theta - V_t) \D t + \gamma \sqrt{V_t} \left(\rho \D W^s_t + \sqrt{1-\rho^2} \D W^v_t\right)
\end{align*}
where $W^s$ and $W^v$ are independent standard Brownian motions.
Then the third moment variation process is represented by
$$ [R,R^2]_t = 2\int_0^t R_s \D [R]_s =  2\int_0^t R_s V_s \D s$$
where we use $\D[R]_s = V_t \D s$.
Note that we do not assume the zero or risk-neutral drift but use the drift $\mu$ under the physical probability and $\mu$ is implicit in the integrand $R_s V_s$.

As in the previous section, consider an investor holding a hedged portfolio composed of an underlying asset $S$ and $\beta S_0$ numbers of the third moment variation swap, i.e., receiving the floating leg $-\beta S_0 [R,R^2]_T$, with maturity $T$.
For simplicity, assume that the fixed leg of the swap is zero.
The log-return of the hedged portfolio over $[0,T]$ is approximated by
$$ X_T = \log \left( \frac{S_T - \beta S_0 [R,R^2]_T}{S_0} \right) \approx R_T - \beta[R,R^2]_T = R_T - 2\beta\int_0^T R_s V_s \D s.$$
Now we explain the backward and forward approaches to compute the distribution of the hedged portfolio.

\subsection{Backward approach}
To compute the probability density function of the portfolio return distribution, first, consider a backward approach based on the Feynman-Kac theorem.
The time $t$ conditional characteristic function of the portfolio's return is
$$ u(r, v, t) =  \E \left[\left.\e^{\I \phi X_T}  \right| R_t = r, V_t = v\right] =\E \left[\left.\exp\left(-2\I\phi\beta \int_t^T R_s V_s \D s \right)\e^{\I\phi R_T} \right|  R_t = r, V_t = v \right].$$
Then, by the theorem, $u(r,v,t)$ satisfies the following partial differential equation (PDE)
\begin{align*}
\frac{\partial u}{\partial t} + \left(\mu-\frac{1}{2}v\right)\frac{\partial u}{\partial r} + \kappa(\theta-v)\frac{\partial u}{\partial v} + \frac{1}{2}v\frac{\partial^2 u}{\partial r^2} + \frac{1}{2}\gamma^2 v\frac{\partial^2 u}{\partial v^2} + \rho \gamma v \frac{\partial^2 u}{\partial r \partial v} = 2\I\phi \beta r v u
\end{align*}
with the terminal condition $u(r,v,T) = \e^{\I\phi r}$.

We can derive the characteristic function with sufficiently enough numbers of grid points $\in [\phi_{\min}, \phi_{\max}]$ by the above PDE with initial points $r=0$ and $v=v_0$,
and compute the probability density function of the hedged portfolio by the discrete Fourier transform.
The drawback of this approach is that the boundary conditions of the PDE is not well-defined
and this cause errors when we apply numerical procedure on the PDE.
Therefore, we apply the forward approach to compute the probability density function.

\subsection{Forward approach}
We have three dimensional stochastic differential equations for the dynamics of the portfolio return:
\[
\left[\begin{array}{c} \D X_t \\ \D R_t \\ \D V_t \end{array}\right] = \left[\begin{array}{c} \mu - \frac{1}{2}V_t  -2\beta R_t V_t \\  \mu - \frac{1}{2}V_t  \\ \kappa(\theta-V_t) \end{array}\right] \D t + 
\left[\begin{array}{ccc} 0 & \sqrt{V_t} & 0 \\ 0 & \sqrt{V_t} & 0 \\ 0 & \gamma \rho \sqrt{V_t} & \gamma \sqrt{1-\rho^2}\sqrt{V_t}  \end{array} \right]
 \left[\begin{array}{c} \D W_t^x \\ \D W_t^s \\ \D W_t^v \end{array} \right]
\]
where $W^x$ is a dummy variable which is not used elsewhere.
The variance-covariance matrix is represented by
\[
\left[\begin{array}{ccc} 0 & \sqrt{V_t} & 0 \\ 0 & \sqrt{V_t} & 0 \\ 0 & \gamma \rho \sqrt{V_t} & \gamma \sqrt{1-\rho^2}\sqrt{V_t}  \end{array} \right] 
\left[\begin{array}{ccc} 0 & 0 & 0 \\ \sqrt{V_t} & \sqrt{V_t} & \gamma \rho \sqrt{V_t} \\ 0 & 0 & \gamma \sqrt{1-\rho^2}\sqrt{V_t}  \end{array} \right]
=\left[\begin{array}{ccc} V_t & V_t & \gamma \rho V_t \\ V_t & V_t & \gamma \rho V_t \\ \gamma\rho V_t &\gamma\rho V_t & \gamma^2 V_t \end{array} \right].
\]
By the forward Kolmogorov equation, also known as the Fokker-Planck equation, we derive the PDE for the joint probability density function $f(x,r,v,t)$ with $x=X_t, r=R_t$ and $v=V_t$ at time $t$ of the three dimensional random vectors $(X_t, R_t, V_t)$:
\begin{align*}
\frac{\partial f}{\partial t} ={}& -\left(\mu -\frac{1}{2}v - 2\beta r v \right)\frac{\partial f}{\partial x} -   \left(\mu - \frac{1}{2}v\right)\frac{\partial f}{\partial r} - \frac{\partial}{\partial v}\kappa(\theta - v)f + \frac{v}{2}\frac{\partial^2 f}{\partial x^2} + \frac{v}{2}\frac{\partial^2 f}{\partial r^2} + \frac{\gamma^2}{2}\frac{\partial^2}{\partial v^2}vf \\
&+ v\frac{\partial^2 f}{\partial x \partial r}
+ \rho\gamma\frac{\partial^2}{\partial x \partial v}vf + \rho\gamma\frac{\partial^2}{\partial r \partial v}vf
\end{align*}
with the initial condition $f(x,r,v,0) = \delta(x)\delta(r)\delta(v-v_0)$ where $\delta$ denotes the Dirac delta function.

\begin{figure}
\centering
\includegraphics[width=8cm]{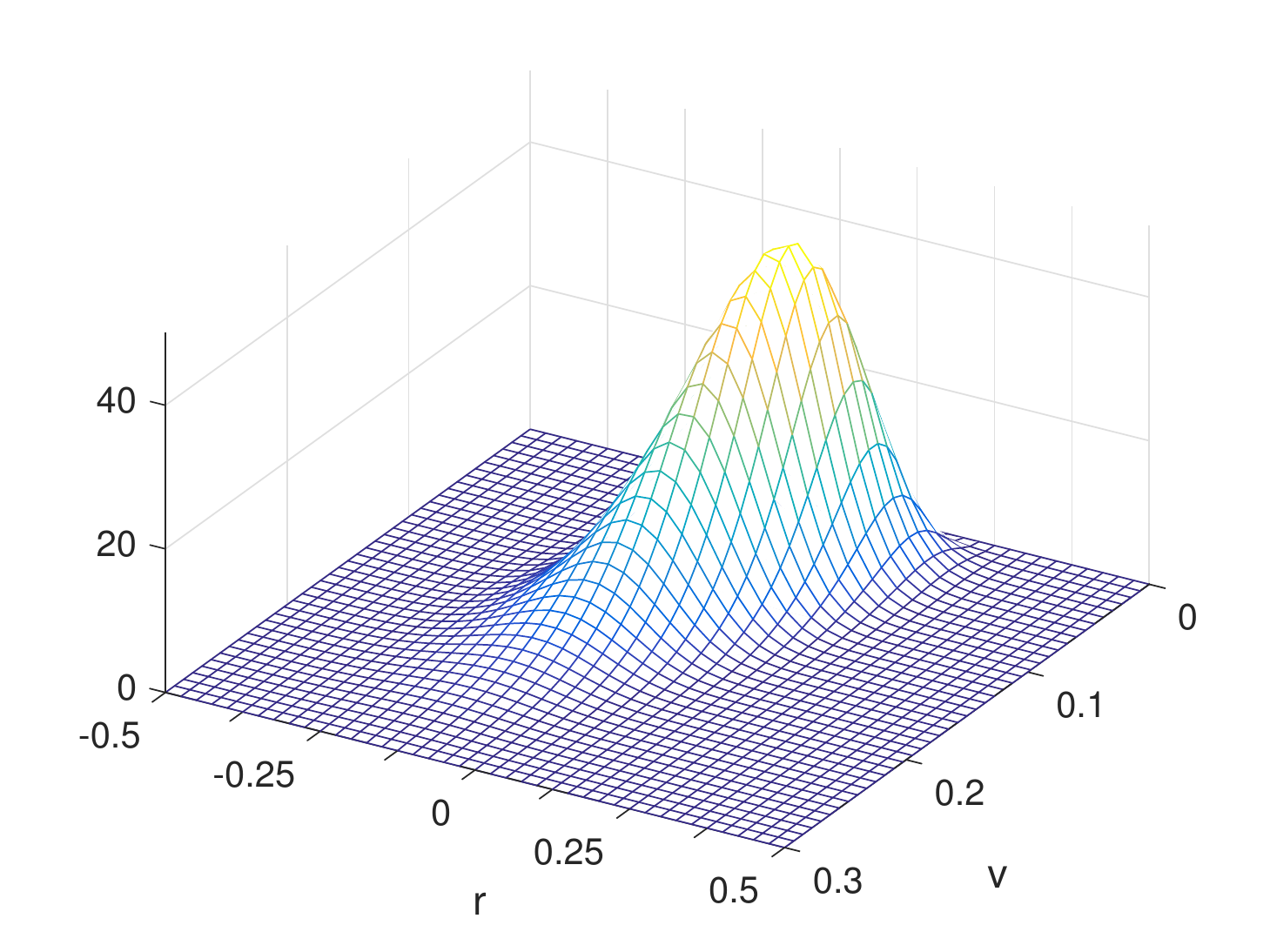}
\caption{When $\phi=0$, the solution of the Eq.~\eqref{Eq:tPDE} is the joint probability density function of the return and variance in Heston's model}\label{Fig:sol0}
\end{figure}

The spatial domain of the PDE is three dimensional.
For the numerical procedure, it is convenient to reduce the dimension of the space.
Since $x$ appears in the PDE only in the derivative operators, we apply the Fourier transform of $f(x,r,v,t)$ with respect to $x$:
$$ \hat f(r,v,t;\phi) = \int_{-\infty}^{\infty} f(x,r,v,t)\e^{-\I\phi x}\D x.$$
The Fourier transforms of the partial derivatives with respect to $x$ are
$$\mathbb F \left \{ \frac{\partial f}{\partial x} \right\} = \I \phi \hat f, \quad \mathbb F \left \{ \frac{\partial^2 f}{\partial x^2} \right\} = - \phi^2 \hat f.$$
Applying the above Fourier transforms to the PDE, we have
\begin{align}
\frac{\partial \hat f}{\partial t} ={}& \left(-\mu + \frac{1}{2}v + \I \phi v + \rho \gamma\right)\frac{\partial \hat f}{\partial r} + \frac{v}{2}\frac{\partial^2 \hat f}{\partial r^2} + \left\{-\kappa(\theta-v) + \gamma^2 + \I \rho \gamma \phi v \right\}\frac{\partial \hat f}{\partial v} + \frac{\gamma^2}{2}v \frac{\partial^2 \hat f}{\partial v^2} \nonumber\\
&+ \rho \gamma v \frac{\partial^2 \hat f}{\partial r \partial v} + \left\{ \I\phi\left( -\mu + \frac{1}{2}v + 2\beta rv \right) - \frac{\phi^2}{2}v + \I \rho \gamma \phi + \kappa \right\} \hat f \label{Eq:tPDE}
\end{align}
with the initial condition $\hat f(r,v,0;\phi) = \delta(r)\delta(v-v_0)$.
When $\phi=0$, the solution of the PDE is reduced to the joint probability density function of $(R_t, V_t)$ under the square root stochastic volatility model as plotted in Figure~\ref{Fig:sol0}.

For brevity, let
\begin{align*}
\mu_r(r,v) &= -\mu + \frac{1}{2}v + \I \phi v + \rho \gamma, \quad \sigma_r(r,v) = \frac{v}{2},\\
\mu_v(r,v) &= -\kappa(\theta-v) + \gamma^2 + \I \rho \gamma \phi v , \quad
\sigma_v(r,v) = \frac{\gamma^2}{2}v, \\
\alpha(r,v) &=  \I\phi\left( -\mu + \frac{1}{2}v + 2\beta rv \right) - \frac{\phi^2}{2}v + \I \rho \gamma \phi + \kappa.
\end{align*}
Then we can rewrite
\begin{align*}
\frac{\partial \hat f}{\partial t} = \mu_r(r,v)\frac{\partial \hat f}{\partial r} + \sigma_r(r,v)\frac{\partial^2 \hat f}{\partial r^2} + \mu_v(r,v)\frac{\partial \hat f}{\partial v} + \sigma_v(r,v) \frac{\partial^2 \hat f}{\partial v^2} + \rho \gamma v \frac{\partial^2 \hat f}{\partial r \partial v} +  \alpha(r,v) \hat f.
\end{align*}

We also construct a PDE for the joint probability density function $g(y,r,v,t)$ of the third moment variation with $y = Y_t := [R^2,R]_t, r=R_t$ and $v=V_t$ at time $t$ of the three dimensional random vectors $(Y_t, R_t, V_t)$:
\begin{align}
\frac{\partial g}{\partial t} = -2 r v \frac{\partial g}{\partial y} - \left(\mu - \frac{1}{2}v\right)\frac{\partial g}{\partial r} - \frac{\partial}{\partial v}\kappa(\theta - v)g  + \frac{v}{2}\frac{\partial^2 g}{\partial r^2} + \frac{\gamma^2}{2}\frac{\partial^2}{\partial v^2}vg + \rho\gamma\frac{\partial^2}{\partial r \partial v}vg \label{Eq:PDE_tm}
\end{align}
A transformed PDE with respect to $y$ is represented by
\begin{align*}
\frac{\partial \hat g}{\partial t} ={}& \left(-\mu +\frac{1}{2}v +\rho \gamma \right)\frac{\partial \hat g}{\partial r} + \frac{v}{2}\frac{\partial^2 \hat g}{\partial r^2} + \{-\kappa(\theta-v)+\gamma^2 \}\frac{\partial \hat g}{\partial v} + \frac{\gamma^2}{2}v\frac{\partial^2 \hat g}{\partial v^2} + \rho\gamma v \frac{\partial^2 \hat g}{\partial r \partial v} \\
&+ (-2\I \phi rv + \kappa)\hat g.
\end{align*}

Now we explain the details of the numerical method to solve the PDE~\eqref{Eq:tPDE}.
The finite difference method for the transformed PDE is employed to compute the numerical solution of the joint distribution.
The spatial domain is restricted to a bounded region $[r_{\min}, r_{\max}] \times [0,v_{\max}]$ 
where $r_{\min} = -r_{\max}$.
The numbers of the grid points of the return $r$ and the volatility $v$ spaces are equal and $N$ denotes the number such that $r_{\min} = r_1 < \cdots < r_N = r_{\max}$ and $0 = v_1 < \cdots < v_N = v_{\max}$.
The difference sizes of the grid points of the return and volatility spaces are denoted by $\Delta r$ and $\Delta v$, respectively.

The derivatives in $r$ and $v$ directions are computed using the central difference scheme when $r_1 < r_i <r_N$ and $v_1 < v_j <v_N$.
Under the scheme, the partial derivatives are approximated by
\begin{align*}
\frac{\partial \hat f}{\partial r}(r_i,v_j) &\approx \frac{\hat f_{i+1,j}-\hat f_{i-1,j}}{2\Delta r} \\
\frac{\partial^2 \hat f}{\partial r^2}(r_i,v_j) &\approx \frac{\hat f_{i+1,j}-2\hat f_{i,j}+ \hat f_{i-1,j}}{(\Delta r)^2}\\
\frac{\partial^2 \hat f}{\partial r\partial v}(r_i,v_j) &\approx \frac{\hat f_{i+1,j+1}-\hat f_{i-1,j+1}- \hat f_{i+1,j-1}+\hat f_{i-1,j-1}}{4\Delta v\Delta r},
\end{align*}
where, for simplicity, the time notation $t$ is omitted and $\hat f_{i,j}$ is an approximation of $\hat f(r_i,v_j)$ under our numerical procedure.
Similarly, the derivatives with respect to $v$ are approximated.
When $r_i$ or $v_j$ has the boundary value of the spatial grid, we use the one-sided difference scheme,
for example,
\begin{align*}
\frac{\partial \hat f}{\partial r}(r_1,v_j) &\approx \frac{\hat f_{2,j}-\hat f_{1,j}}{\Delta r} \\
\frac{\partial^2 \hat f}{\partial r^2}(r_1,v_j) &\approx \frac{\hat f_{3,j}-2\hat f_{2,j}+ \hat f_{1,j}}{(\Delta r)^2}.
\end{align*}

For the time discretization, the alternating direction implicit (ADI) method of \cite{PeacemanRachford} is used.
The direction of $r$ is first treated implicitly and the next, the direction of $v$ is treated implicitly.
The error due to the explicit scheme is reduced by the decreased error in the next implicit step.
The mixed derivative term is calculated explicitly.
The time points are distributed equally with the difference $\Delta t$.
Between two discrete time points $t_n$ and $t_{n+1}$, there is an intermediate point $t_{n+1/2}$.

To apply the ADI scheme, we have a finite difference formula for the intermediate step.
For $1 < i < N$,
\begin{align*}
\frac{\hat f^{n+1/2}_{i,j} - \hat f^{n}_{i,j}}{\Delta t/2} ={}& \mu_r(r_i,v_j)\frac{\hat f^{n+1/2}_{i+1,j}-\hat f^{n+1/2}_{i-1,j}}{2\Delta r} + \sigma_r(r_i,v_j)\frac{\hat f^{n+1/2}_{i+1,j} - 2 \hat f^{n+1/2}_{i,j} + \hat f^{n+1/2}_{i-1,j}}{\Delta r^2} \\
&+ \mu_v(r_i,v_j) \frac{\hat f^{n}_{i,j+1}- \hat f^{n}_{i,j-1}}{2 \Delta v} + \sigma_v(r_i,v_j)\frac{\hat f^{n}_{i,j+1} -2\hat f^{n}_{i,j} + \hat f^{n}_{i,j-1}}{(\Delta v)^2} \\
&+ \rho \gamma v_j \frac{\hat f^{n}_{i+1,j+1} - \hat f^{n}_{i-1,j+1} - \hat f^{n}_{i+1,j-1} + \hat f^{n}_{i-1, j-1}}{4\Delta r\Delta v} + \alpha(r_i,v_j) \hat f^{n+1/2}_{i,j}.
\end{align*}
We rewrite
\begin{align}
\left(\frac{\mu_r}{2\Delta r} - \frac{\sigma_r}{(\Delta r)^2}\right)\hat f^{n+1/2}_{i-1,j} + \left(\frac{2}{\Delta t} + \frac{2\sigma_r}{(\Delta r)^2}  -\alpha\right)\hat f^{n+1/2}_{i,j} + \left(-\frac{\mu_r}{2\Delta r} - \frac{\sigma_r}{(\Delta r)^2} \right)\hat f^{n+1/2}_{i+1,j} = b_{i,j} \label{Eq:diagonal1}
\end{align}
where
\begin{align*}
b_{i,j} ={}& \frac{2\hat f^n_{i,j}}{\Delta t} + \mu_v \frac{\hat f^{n}_{i,j+1}- \hat f^{n}_{i,j-1}}{2 \Delta v} + \sigma_v\frac{\hat f^{n}_{i,j+1} -2\hat f^{n}_{i,j} + \hat f^{n}_{i,j-1}}{(\Delta v)^2} \\
&+ \rho \gamma v_j \frac{\hat f^{n}_{i+1,j+1} - \hat f^{n}_{i-1,j+1} - \hat f^{n}_{i+1,j-1} + \hat f^{n}_{i-1, j-1}}{4\Delta r\Delta v}
\end{align*}
and without confusion, let $\mu_r = \mu_r(r_i,v_j)$ and similarly for $\mu_v$, $\sigma_r$, $\sigma_v$ and $\alpha$.

When $i=1$, one-sided difference schemes are used and we have
\begin{align*}
\frac{\hat f^{n+1/2}_{1,j} - \hat f^{n}_{1,j}}{\Delta t/2} ={}& \mu_r\frac{\hat f^{n+1/2}_{2,j} - \hat f^{n+1/2}_{1,j}}{\Delta r} + \sigma_r \frac{\hat f^{n+1/2}_{3,j} - 2\hat f^{n+1/2}_{2,j}+\hat f^{n+1/2}_{1,j}}{(\Delta r)^2} \\
&+ \mu_v \frac{\hat f^{n}_{1,j+1}- \hat f^{n}_{1,j-1}}{2 \Delta v} + \sigma_v \frac{\hat f^{n}_{1,j+1} -2\hat f^{n}_{1,j} + \hat f^{n}_{1,j-1}}{(\Delta v)^2} \\
&+ \rho \gamma v_j \frac{\hat f^{n}_{2,j+1} - \hat f^{n}_{1,j+1} - \hat f^{n}_{2,j-1} + \hat f^{n}_{1, j-1}}{2\Delta r\Delta v} + \alpha \hat f^{n+1/2}_{1,j}
\end{align*}
and
\begin{align}
\left( \frac{2}{\Delta t} + \frac{\mu_r}{\Delta r} - \frac{\sigma_r}{(\Delta r)^2} - \alpha\right) \hat f^{n+1/2}_{1,j} + \left( -\frac{\mu_r}{\Delta r} + \frac{2\sigma_r}{(\Delta r)^2} \right)\hat f^{n+1/2}_{2,j} - \frac{\sigma_r}{ (\Delta r)^2}\hat f^{n+1/2}_{3,j}  = b_{1,j}.\label{Eq:firstrow1}
\end{align}
where
\begin{align*}
b_{1,j} ={}& \frac{2\hat f^n_{1,j}}{\Delta t} + \mu_v \frac{\hat f^{n}_{1,j+1}- \hat f^{n}_{1,j-1}}{2 \Delta v} + \sigma_v\frac{\hat f^{n}_{1,j+1} -2\hat f^{n}_{1,j} + \hat f^{n}_{1,j-1}}{(\Delta v)^2} \\
&+ \rho \gamma v_j \frac{\hat f^{n}_{2,j+1} - \hat f^{n}_{1,j+1} - \hat f^{n}_{2,j-1} + \hat f^{n}_{1, j-1}}{2\Delta r\Delta v}.
\end{align*}
Similarly, when $i=N$, we have 
\begin{align}
-\frac{\sigma_r}{(\Delta r)^2} \hat f^{n+1/2}_{N-2,j} + \left( \frac{\mu_r}{\Delta r} + \frac{2\sigma_v}{(\Delta r)^2} \right) \hat f^{n+1/2}_{N-1,j} + \left( \frac{2}{\Delta t} - \frac{\mu_r}{\Delta r} - \frac{\sigma_v}{(\Delta r)^2} - \alpha \right) \hat f^{n+1/2}_{N,j} = b_{N,j} \label{Eq:lastrow1}
\end{align}
where
\begin{align*}
b_{N,j} ={}& \frac{2\hat f^n_{N,j}}{\Delta t} + \mu_v \frac{\hat f^{n}_{N,j+1}- \hat f^{n}_{N,j-1}}{2 \Delta v} + \sigma_v\frac{\hat f^{n}_{N,j+1} -2\hat f^{n}_{N,j} + \hat f^{n}_{N,j-1}}{(\Delta v)^2} \\
&+ \rho \gamma v_j \frac{\hat f^{n}_{N,j+1} - \hat f^{n}_{N-1,j+1} - \hat f^{n}_{N,j-1} + \hat f^{n}_{N-1, j-1}}{2\Delta r\Delta v}.
\end{align*}

Let $\hat{\mathbf f}^{n+1/2}_j$ and $\mathbf b_j$ denote the $j$-th column vectors that consist of $\hat f^{n+1/2}_{i,j}$ and $b_{i,j}$, respectively.
Then, for each $j$, we have a matrix multiplication form
\begin{equation}
A^{(j)}_r \hat{\mathbf f}^{n+1/2}_j = \mathbf b_j \label{Eq:matrix_eq}
\end{equation}
where $A_r^{(j)}$ is a tridiagonal matrix except the first and last row:
\[ A^{(j)}_r = 
\left( 
\begin{array}{cccccc}
a^{(r,j)}_{1,1} &  a^{(r,j)}_{1,2} & a^{(r,j)}_{1,3} & \cdots & 0 &0 \\
a^{(r,j)}_{2,1} & a^{(r,j)}_{2,2} & a^{(r,j)}_{2,3} & 0  &  & 0 \\
0 & \ddots & \ddots & \ddots & 0 & \vdots   \\
\vdots   & 0 & \ddots & \ddots & \ddots & 0  \\
0  &        & 0 & a^{(r,j)}_{N-1,N-2} & a^{(r,j)}_{N-1,N-1} &a^{(r,j)}_{N-1,N} \\
0 & 0 & \cdots & a^{(r,j)}_{N,N-2} & a^{(r,j)}_{N,N-1}   & a^{(r,j)}_{N,N}
\end{array} 
\right)
\]
where the entries are determined by Eqs.~\eqref{Eq:diagonal1},\eqref{Eq:firstrow1} and \eqref{Eq:lastrow1}.
For example, if $1<1<N$, then, by Eq.~\eqref{Eq:diagonal1}, we have
\begin{align*}
a^{(r,j)}_{i,i} &= \frac{2}{\Delta t} + \frac{2\sigma_r(r_i, v_j)}{(\Delta r)^2}  -\alpha(r_i, v_j) \\
a^{(r,j)}_{i, i-1} &= \frac{\mu_r(r_i, v_j)}{2\Delta r} - \frac{\sigma_r(r_i, v_j)}{(\Delta r)^2}\\
a^{(r,j)}_{i, i+1} &= -\frac{\mu_r(r_i, v_j)}{2\Delta r} - \frac{\sigma_r(r_i, v_j)}{(\Delta r)^2}. 
\end{align*}
The matrix is sparse and the solution of the Eq.~\eqref{Eq:matrix_eq} can be solved efficiently.

For the next step, we apply the finite difference scheme implicitly on $v$-direction. 
For $1 < j < N$, we have
\begin{align*}
\frac{\hat f^{n+1}_{i,j} - \hat f^{n+1/2}_{i,j}}{\Delta t/2} ={}& \mu_r\frac{\hat f^{n+1/2}_{i+1,j}-\hat f^{n+1/2}_{i-1,j}}{2\Delta r} + \sigma_v \frac{\hat f^{n+1/2}_{i+1,j} - 2 \hat f^{n+1/2}_{i,j} + \hat f^{n+1/2}_{i-1,j}}{\Delta r^2} \\
&+ \mu_v \frac{\hat f^{n+1}_{i,j+1}- \hat f^{n+1}_{i,j-1}}{2 \Delta v} + \sigma_v\frac{\hat f^{n+1}_{i,j+1} -2\hat f^{n+1}_{i,j} + \hat f^{n+1}_{i,j-1}}{(\Delta v)^2} \\
&+ \rho \gamma v_j \frac{\hat f^{n+1/2}_{i+1,j+1} - \hat f^{n+1/2}_{i-1,j+1} - \hat f^{n+1/2}_{i+1,j-1} + \hat f^{n+1/2}_{i-1, j-1}}{4\Delta r\Delta v} + \alpha \hat f^{n+1}_{i,j}.
\end{align*}
We rewrite
\begin{align}
\left(\frac{\mu_v}{2\Delta v} - \frac{\sigma_v}{(\Delta v)^2}\right) \hat f^{n+1}_{i,j-1} + \left(\frac{2}{\Delta t} + \frac{2\sigma_v}{(\Delta v)^2} -\alpha\right) \hat f^{n+1}_{i,j} + \left(-\frac{\mu_v}{2\Delta v} - \frac{\sigma_v}{(\Delta v)^2}\right) \hat f^{n+1}_{i,j+1} = c_{i,j}\label{Eq:diagonal2}
\end{align}
where
\begin{align*}
c_{i,j}={}&\frac{2\hat f^{n+1/2}_{i,j}}{\Delta t} + \mu_v\frac{\hat f^{n+1/2}_{i+1,j}-\hat f^{n+1/2}_{i-1,j}}{2\Delta r} + \sigma_v\frac{\hat f^{n+1/2}_{i+1,j} - 2 \hat f^{n+1/2}_{i,j} + \hat f^{n+1/2}_{i-1,j}}{\Delta r^2} \\
&+ \rho \gamma v_j \frac{\hat f^{n+1/2}_{i+1,j+1} - \hat f^{n+1/2}_{i-1,j+1} - \hat f^{n+1/2}_{i+1,j-1} + \hat f^{n+1/2}_{i-1, j-1}}{4\Delta r\Delta v}.
\end{align*}
Similarly with the previous step, when $j=1$ or $j=N$, we use the one-sided different schemes and we have
\begin{align}
\left(\frac{2}{\Delta t} + \frac{\mu_v}{\Delta v} - \frac{\sigma_v}{(\Delta v)^2} - \alpha \right) \hat f_{i,1}^n + \left( -\frac{\mu_v}{\Delta v} + \frac{2\sigma_v}{(\Delta v)^2}v_j \right)\hat f_{i,2}^n - \frac{\sigma_v }{(\Delta v)^2} \hat f_{i,3}^n = c_{i,1}\label{Eq:firstrow2}
\end{align}
and
\begin{align}
-\frac{\sigma_v }{(\Delta v)^2}\hat f_{i,N-2}^n + \left( \frac{\mu_v}{\Delta v} + \frac{2\sigma_v}{(\Delta v)^2} \right)\hat f_{i,N-1}^n + \left( \frac{2}{\Delta t} - \frac{\mu_v}{\Delta v} - \frac{\sigma_v}{(\Delta v)^2}v_j - \alpha \right)\hat f_{i,N}^n = c_{i,N}\label{Eq:lastrow2}
\end{align}
where $c_{i,1}$ and $c_{i,N}$ are defined by the one-sided schemes as in the previous step.
Thus, for each $i$, we have a matrix multiplication form with raw vectors of $\hat{\mathbf{f}}_i^{n}$ and $\mathbf{c}^{n}_i$
$$ A_v^{(i)} \left( \hat{\mathbf{f}}_i^{n} \right)^\mathsf{T} = \left( \mathbf{c}^n_i \right)^\mathsf{T} $$
where $^\mathsf{T}$ denotes the non-conjugate transpose and
\[ 
A_v^{(i)} = \left( 
\begin{array}{cccccc}
a^{(v,i)}_{1,1} &  a^{(v,i)}_{1,2} & a^{(v,i)}_{1,3} & \cdots & 0 &0 \\
a^{(v,i)}_{2,1} & a^{(v,i)}_{2,2} & a^{(v,i)}_{2,3} & 0  &  & 0 \\
0 & \ddots & \ddots & \ddots & 0 & \vdots   \\
\vdots   & 0 & \ddots & \ddots & \ddots & 0  \\
0  &        & 0 & a^{(v,i)}_{N-1,N-2} & a^{(v,i)}_{N-1,N-1} &a^{(v,i)}_{N-1,N} \\
0 & 0 & \cdots & a^{(v,i)}_{N,N-2} & a^{(v,i)}_{N,N-1}   & a^{(v,i)}_{N,N}
\end{array} 
\right)
\]
where the entries are determined by Eqs~\eqref{Eq:diagonal2},\eqref{Eq:firstrow2} and \eqref{Eq:lastrow2}.

The boundary conditions are imposed as
$$ \hat f(r_{\min},v,t;\phi) = \hat f(r_{\max},v,t;\phi) = \hat f(r,0,t;\phi) = \hat f(r,v_{\max},t;\phi) = 0$$
under the assumption that the parameters in the model satisfy the Feller condition to guarantee the positiveness of the variance process: $2\kappa\theta > \gamma^2$.

\subsection{Probability density function}

\begin{figure}
\centering
\includegraphics[width=0.45\textwidth]{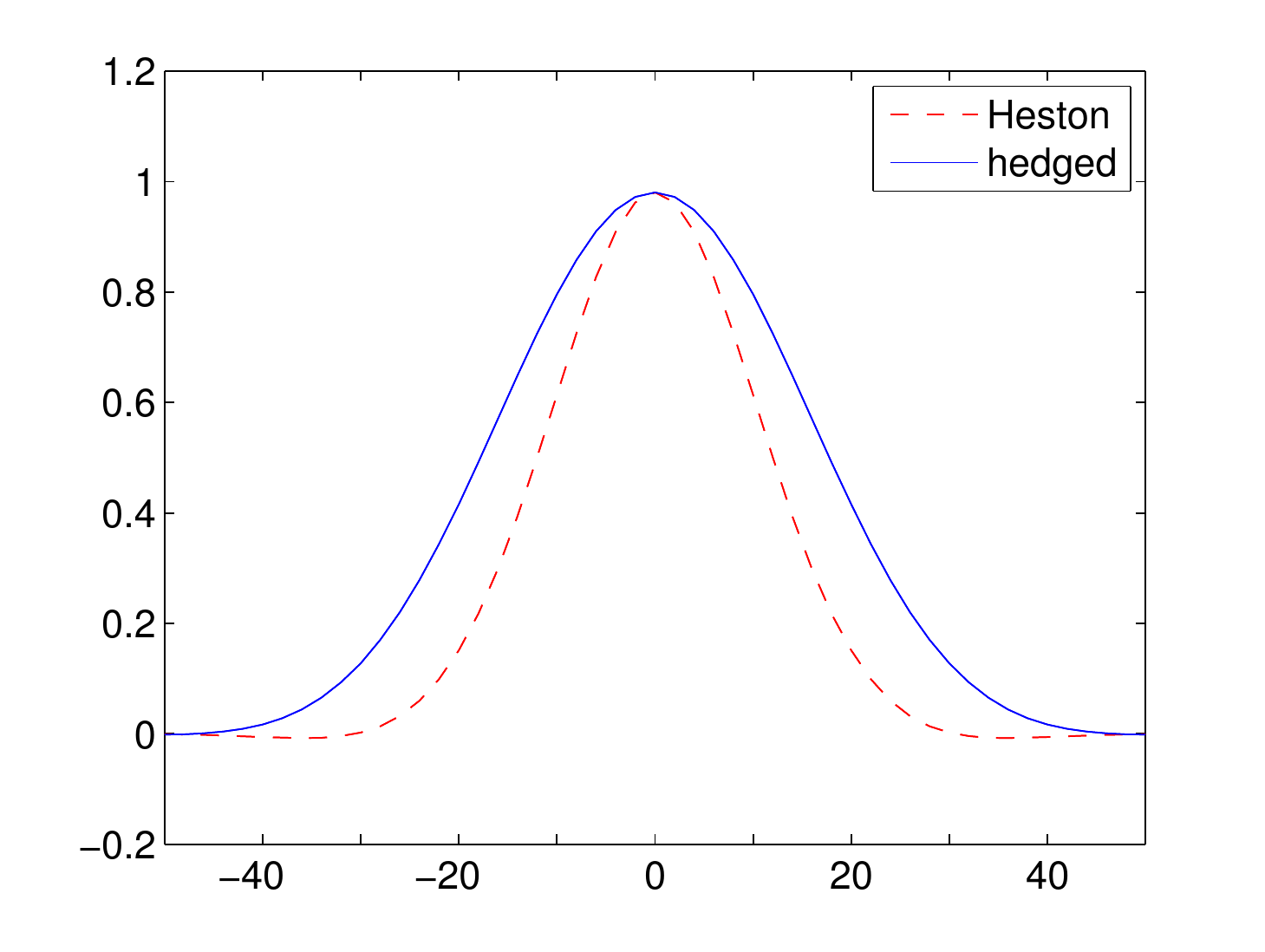}
\includegraphics[width=0.45\textwidth]{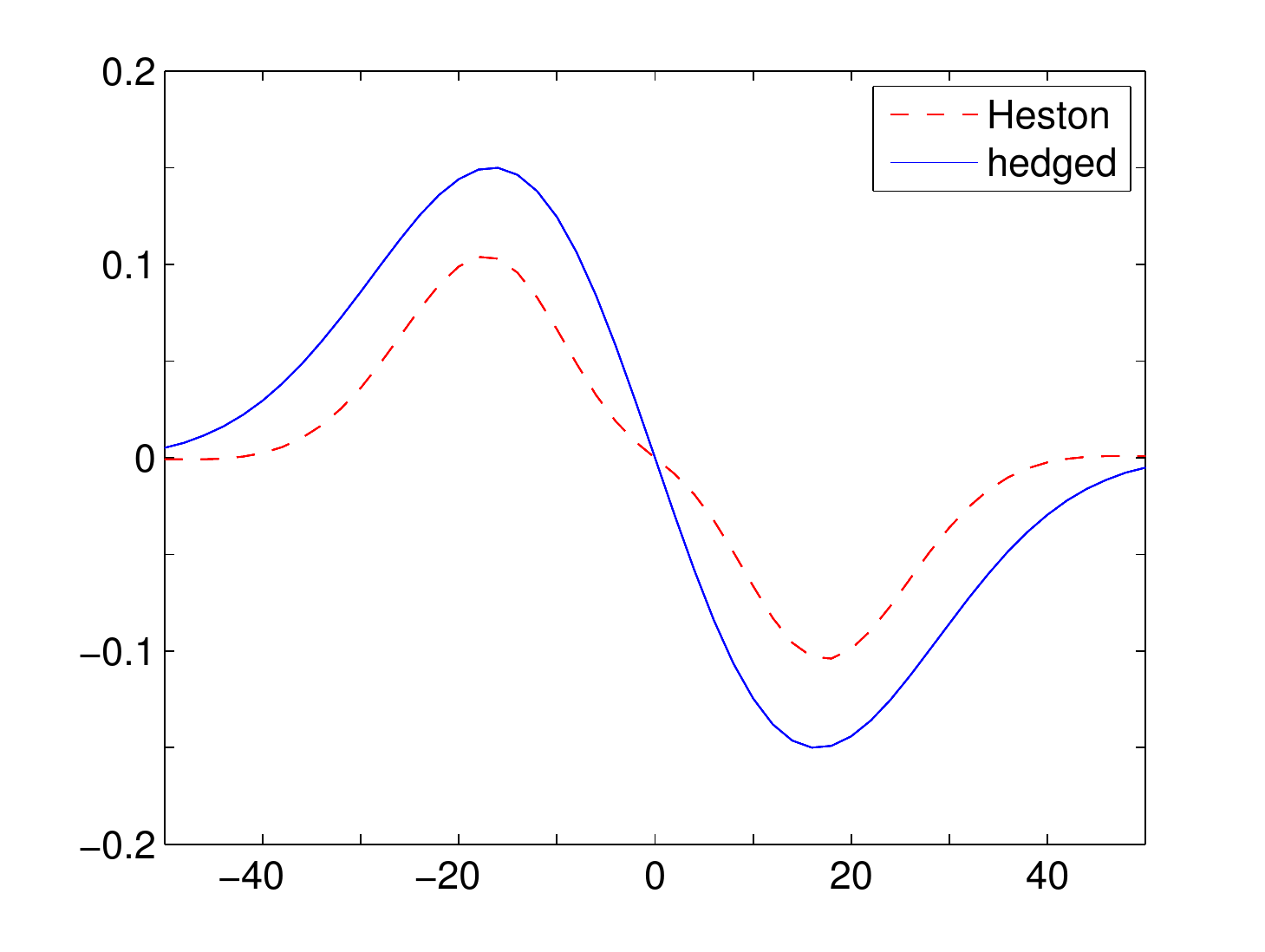}
\caption{The real (left) and imaginary (right) parts of the characteristic functions with $\beta = 0$ (dashed) and 40 (solid)}\label{Fig:cf}
\end{figure}

By performing the numerical procedure, we get the characteristic function of $X_T$ as plotted in Figure~\ref{Fig:cf}
with parameter settings $\kappa = 18, \theta = 0.1, \gamma = 1, \rho = -0.62$ and $T = 0.1$.
In the figure, the real (left) and imaginary (right) parts of the characteristic functions are presented for underlying asset (dashed) and hedged portfolio with $\beta = 40$ (solid).
Taking the transform to the characteristic functions, we compute the joint probability density function of $(X_T, R_T, V_T)$.

By applying the ADI scheme, the numerical procedure takes much less time compared to the explicit scheme
as the time step needed to ensure the numerical stability is much larger than in the case of the explicit scheme.
In the ADI scheme, with a spatial grid of $[R_{\min}, R_{\max}] = [-0.5,0.5]$, $[V_{\min}, V_{\max}] = [0, 0.3]$, $\Delta r = 0.025$ and $\Delta v =  0.0075$, we a stable result with $\Delta t = 0.001$.
Meanwhile, for the explicit method, $\Delta t$ needs be around $2 \times 10^{-4}$ when the same spatial grid is used.

\begin{figure}
\centering
\includegraphics[width=0.45\textwidth]{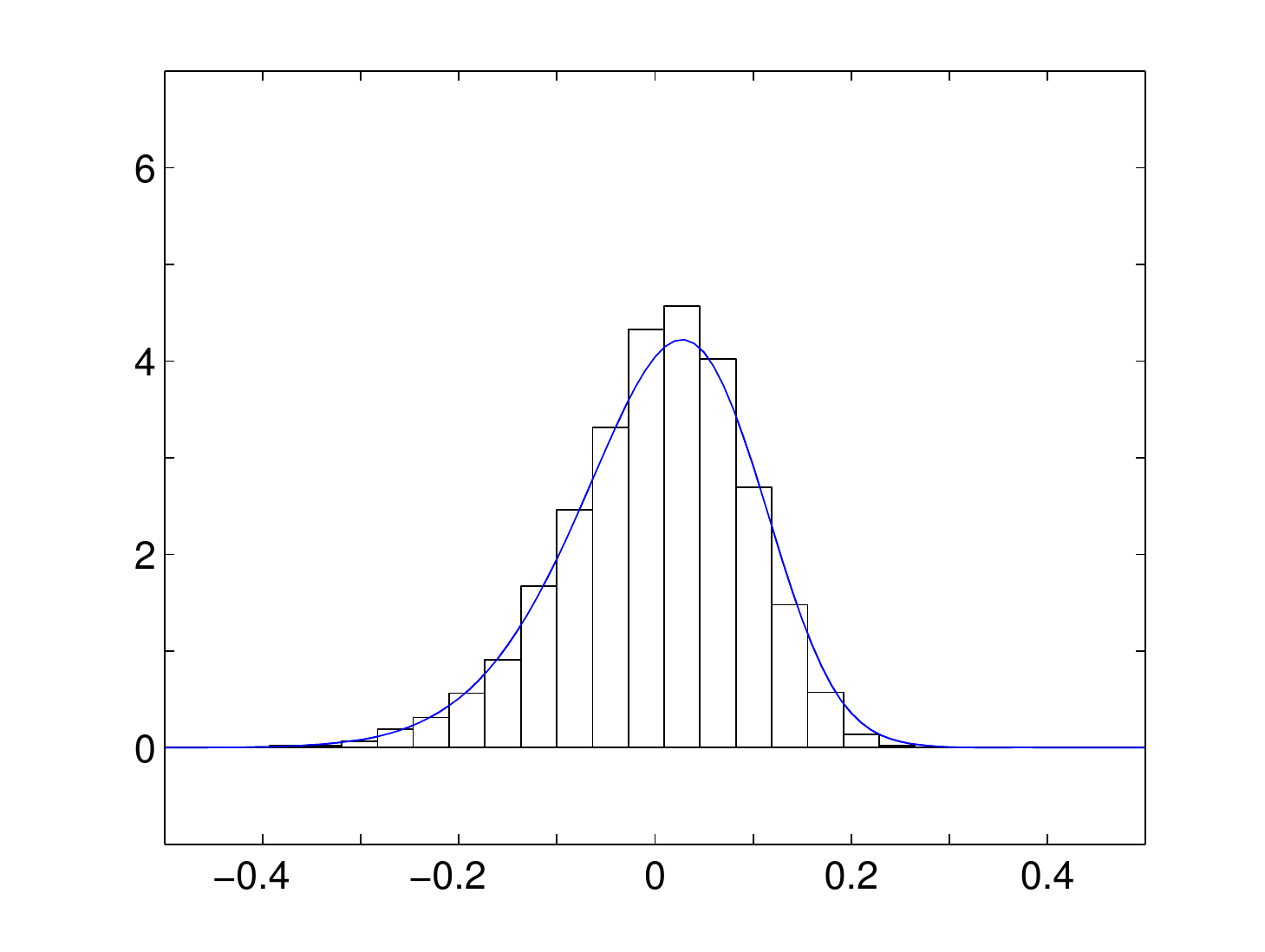}
\includegraphics[width=0.45\textwidth]{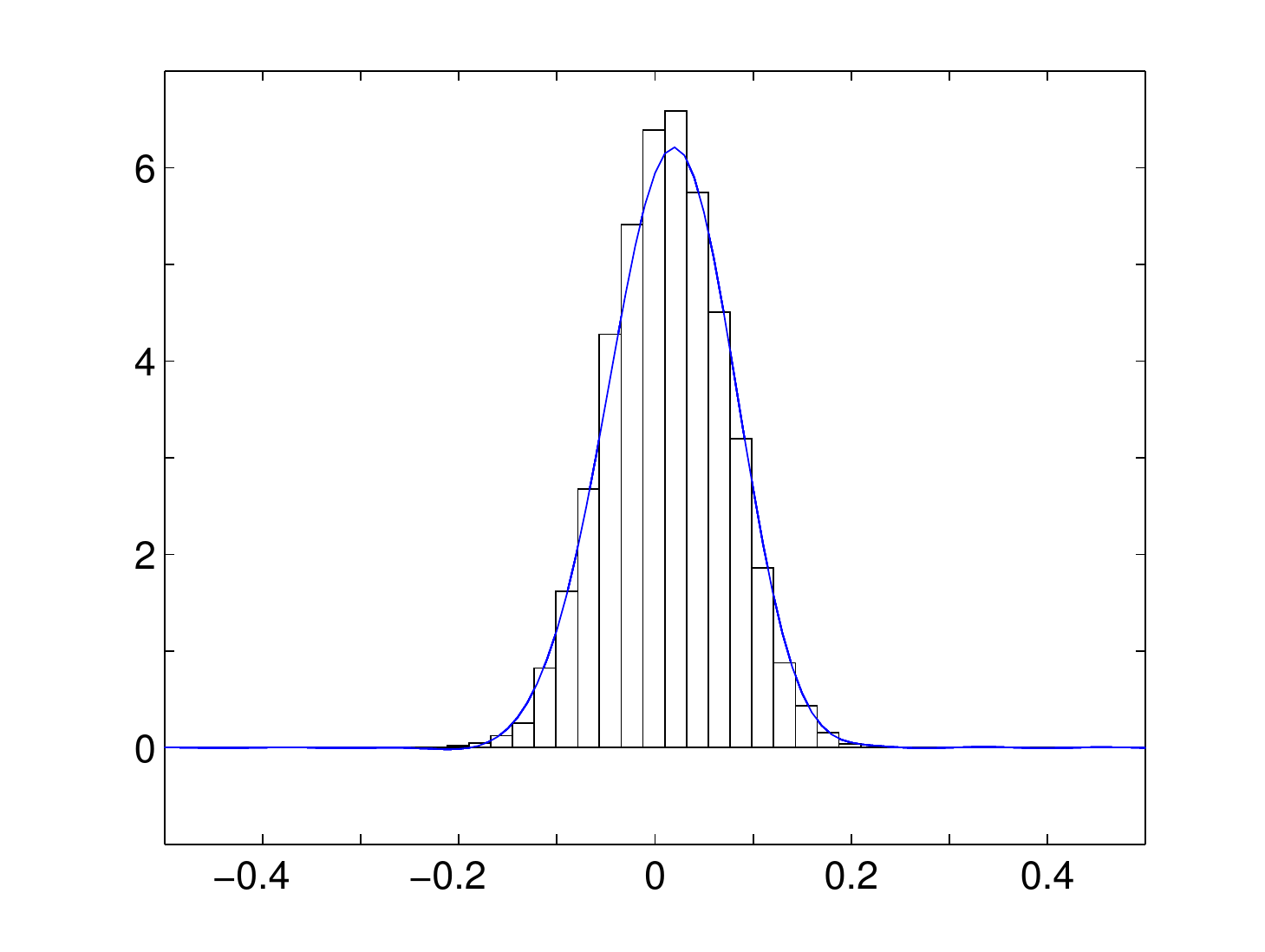}
\caption{Probability density functions and histograms of simulated data : underlying asset (left) and hedged portfolio (right)}\label{Fig:hist}
\end{figure}

\begin{figure}
        \centering
        \begin{subfigure}[b]{0.45\textwidth}
                \includegraphics[width=\textwidth]{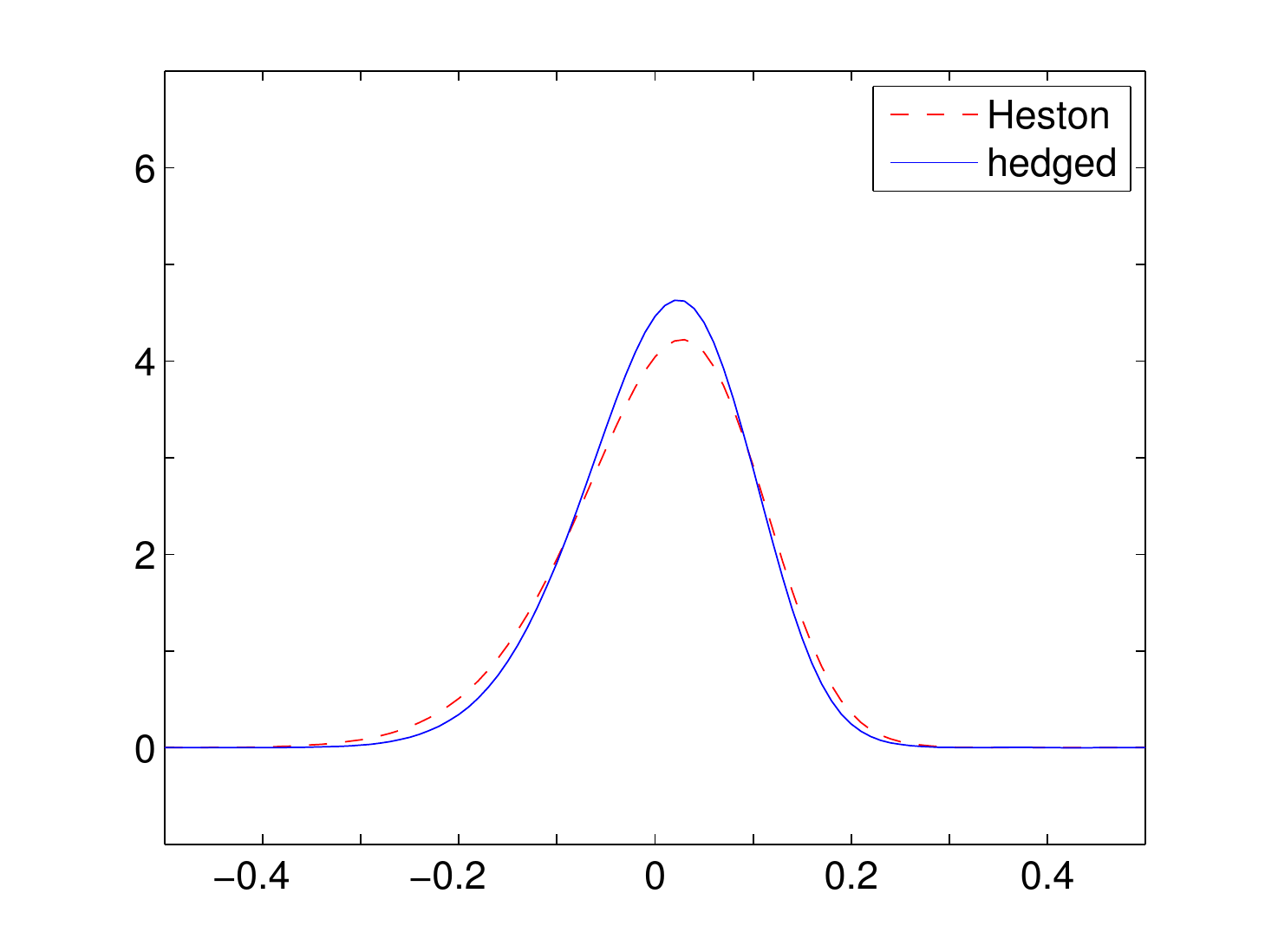}
                \caption{$\beta = 10$}
                \label{Fig:pdf_beta10}
        \end{subfigure}
  		\quad
        \begin{subfigure}[b]{0.45\textwidth}
                \includegraphics[width=\textwidth]{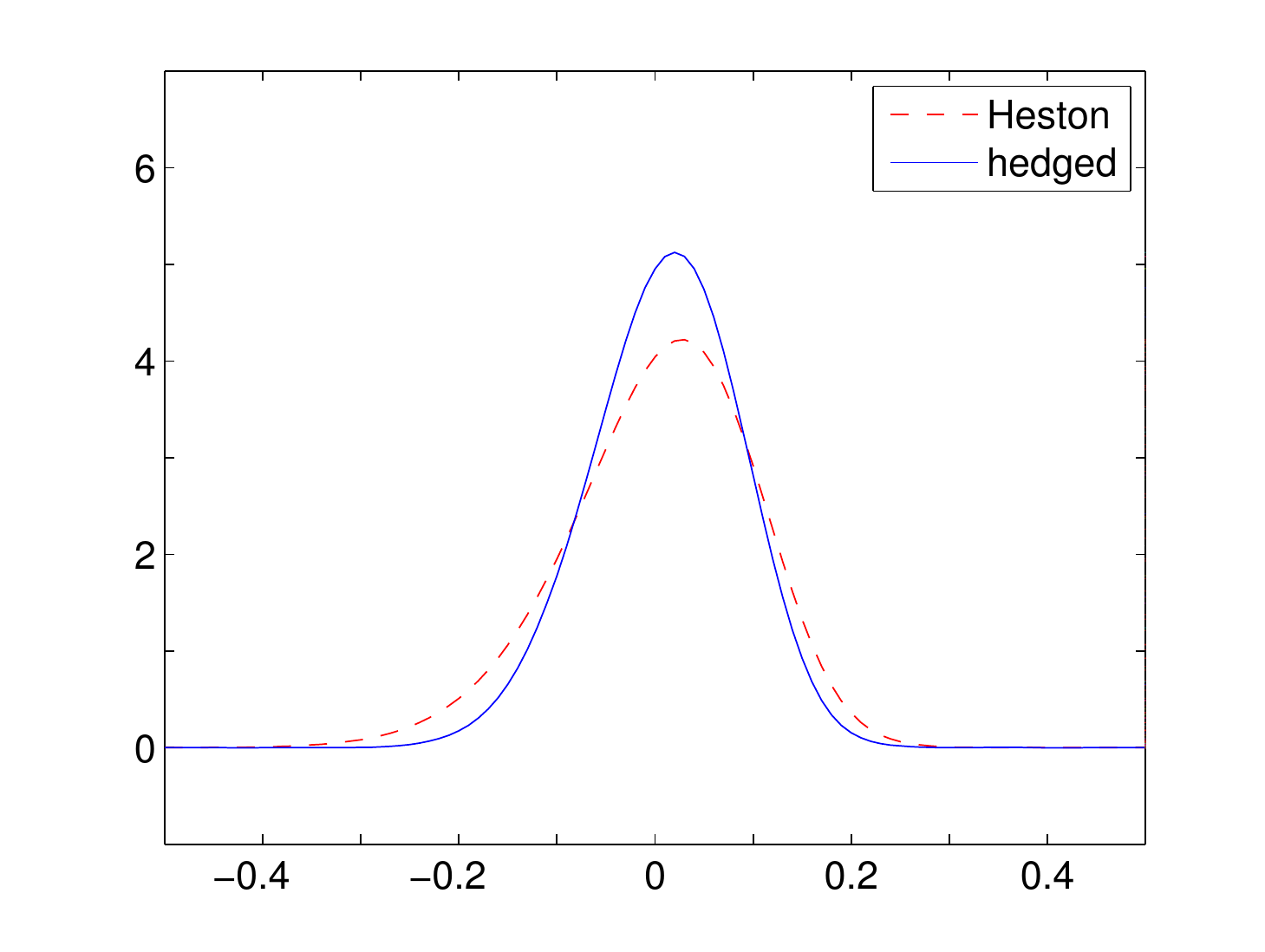}
                \caption{$\beta = 20$}
                \label{Fig:pdf_beta20}
        \end{subfigure}
        
        \begin{subfigure}[b]{0.45\textwidth}
                \includegraphics[width=\textwidth]{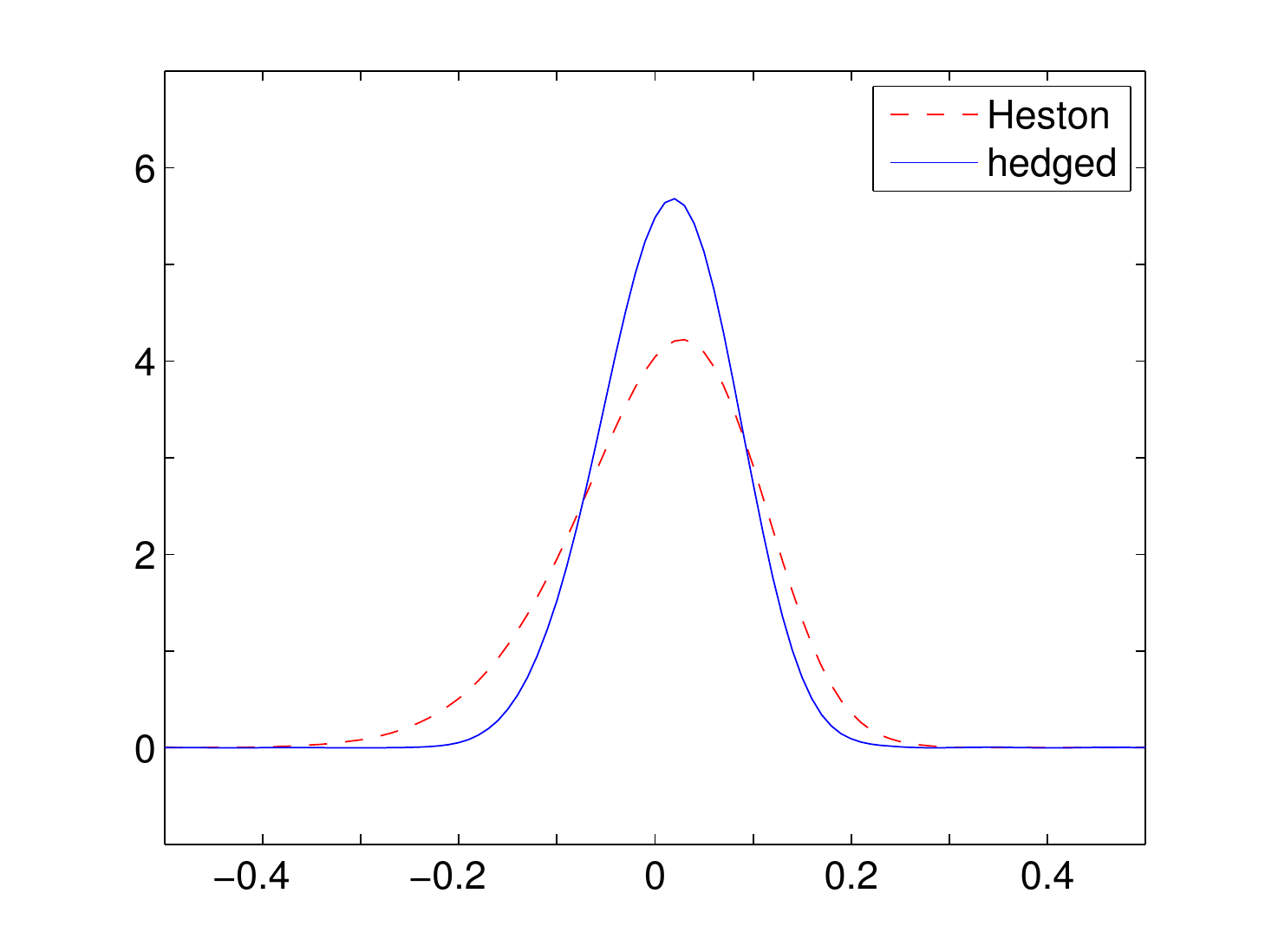}
                \caption{$\beta = 30$}
                \label{Fig:pdf_beta30}
        \end{subfigure}
  		\quad
		\begin{subfigure}[b]{0.45\textwidth}
                \includegraphics[width=\textwidth]{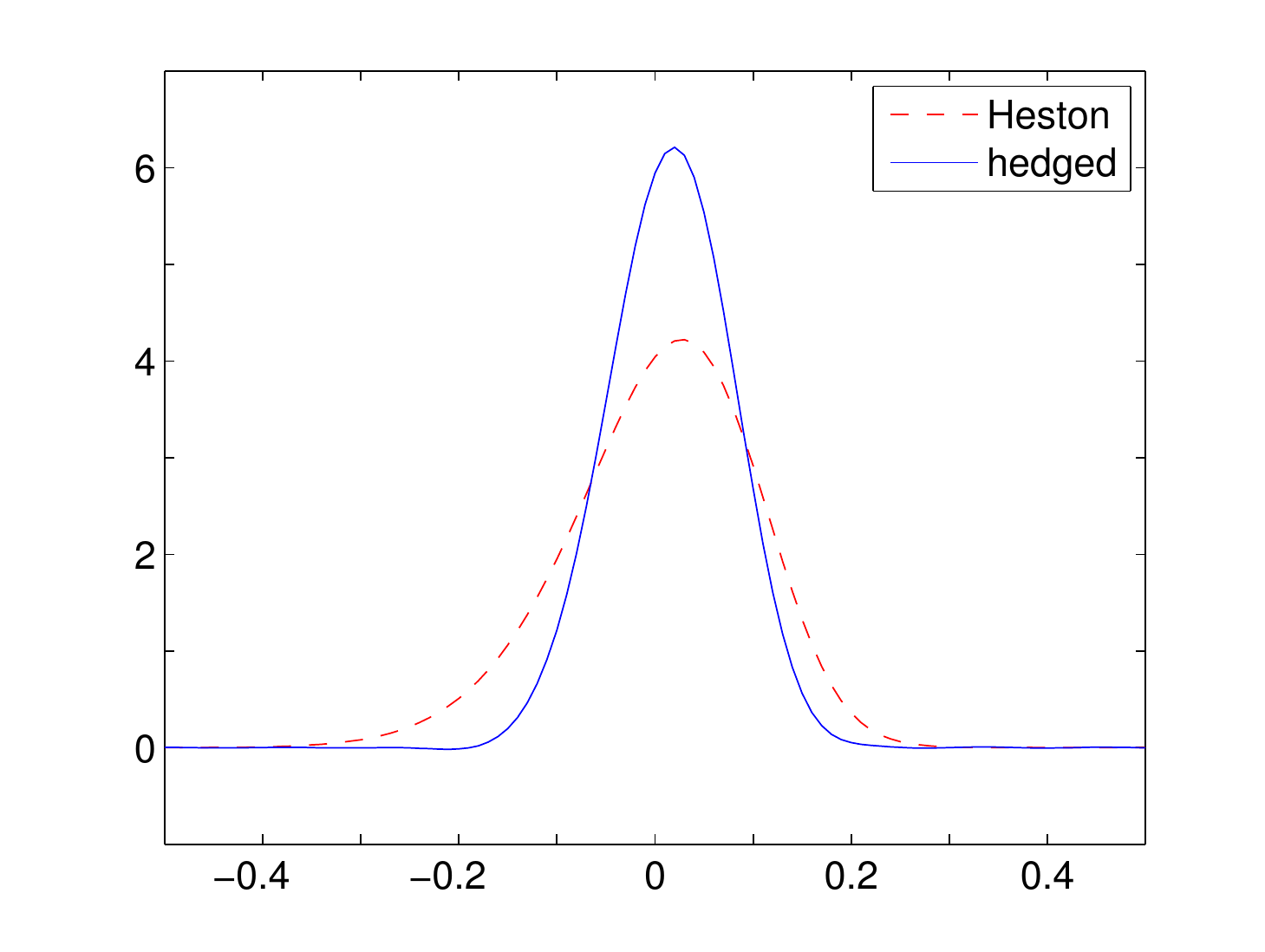}
                \caption{$\beta = 40$}
                \label{Fig:pdf_beta40}
        \end{subfigure}
        
		\begin{subfigure}[b]{0.45\textwidth}
                \includegraphics[width=\textwidth]{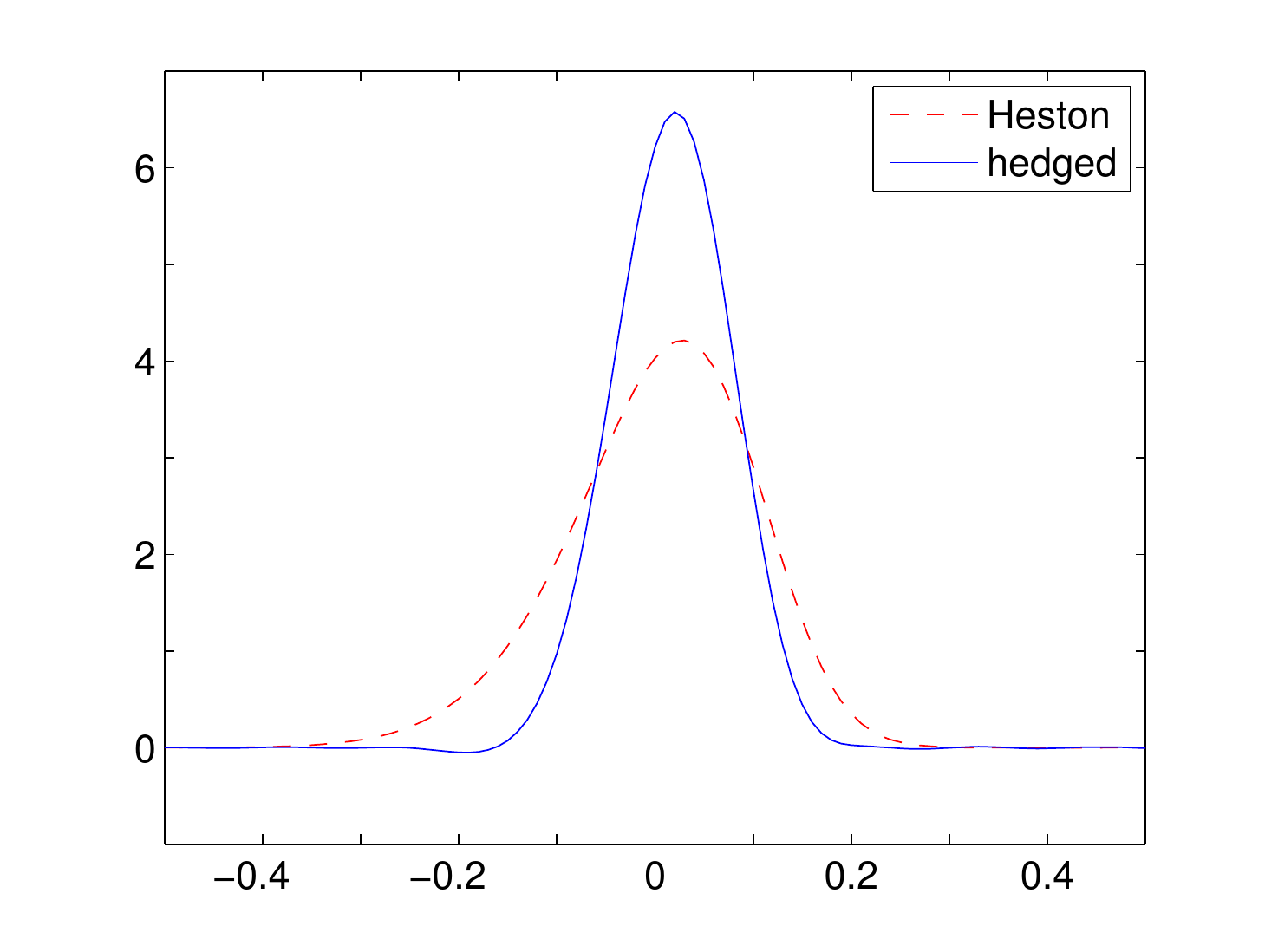}
                \caption{$\beta = 50$}
                \label{Fig:pdf_beta50}
        \end{subfigure}
  		\quad
		\begin{subfigure}[b]{0.45\textwidth}
                \includegraphics[width=\textwidth]{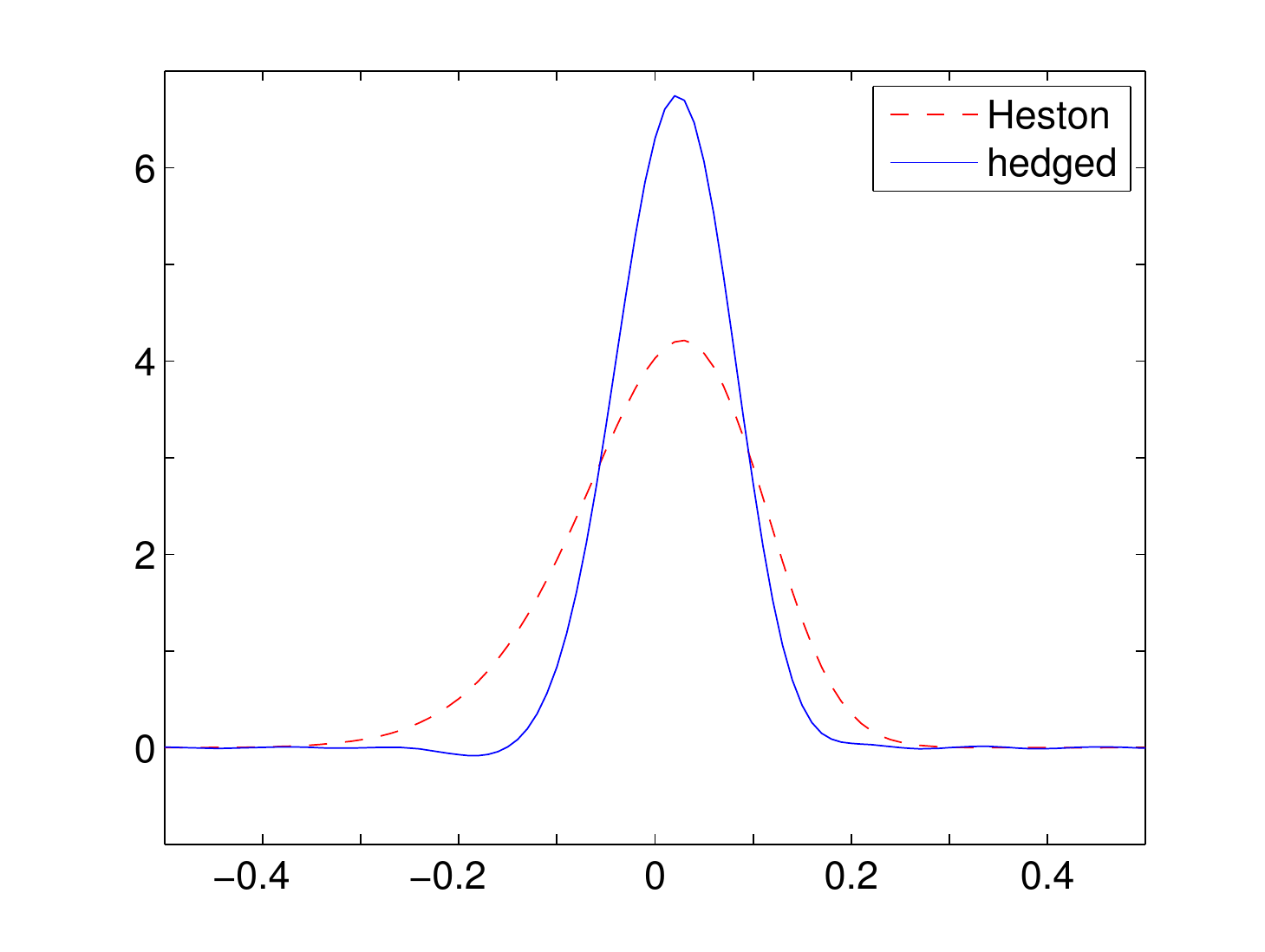}
                \caption{$\beta = 60$}
                \label{Fig:pdf_beta60}
        \end{subfigure}
		\caption{Probability density functions with various hedge numbers $\beta$ (solid) compared to the underlying asset under Heston's model (dashed)}\label{Fig:pdf}
\end{figure}

Figure~\ref{Fig:hist} shows the probability density functions of the underlying asset (left) and the hedged portfolio (right) with hedge number $\beta = 40$ and $T = 0.1$ compared to the histograms of simulated data.
The sample size of the simulation is $10^5$.
Figure~\ref{Fig:pdf} presents the probability density functions of the hedged portfolio returns (solid) compared to the return of the underlying asset (dashed) with various hedge numbers $\beta = 10,20,30,40,50$ and $60$.
The hedged portfolios have more Gaussian-like thin-tail distribution compared to the distribution of the underlying asset.

Table~\ref{Table:statistics} lists the numerically computed mean, standard deviation, skewness and kurtosis of the return distributions of the portfolios with various hedge numbers $\beta = 0, 10, \ldots, 60$.
The table suggests that with $\beta$ between 30 and 40, the skewness of the portfolio return is around zero and has minimal kurtosis.
This result is consistent with the simulation study where the optimal hedge number is reported to be $38.42$.
A simulation study is performed with the same method for the empirical analysis explained in Section~\ref{Sect:Empirical}.

\begin{figure}
\centering
\includegraphics[width=0.45\textwidth]{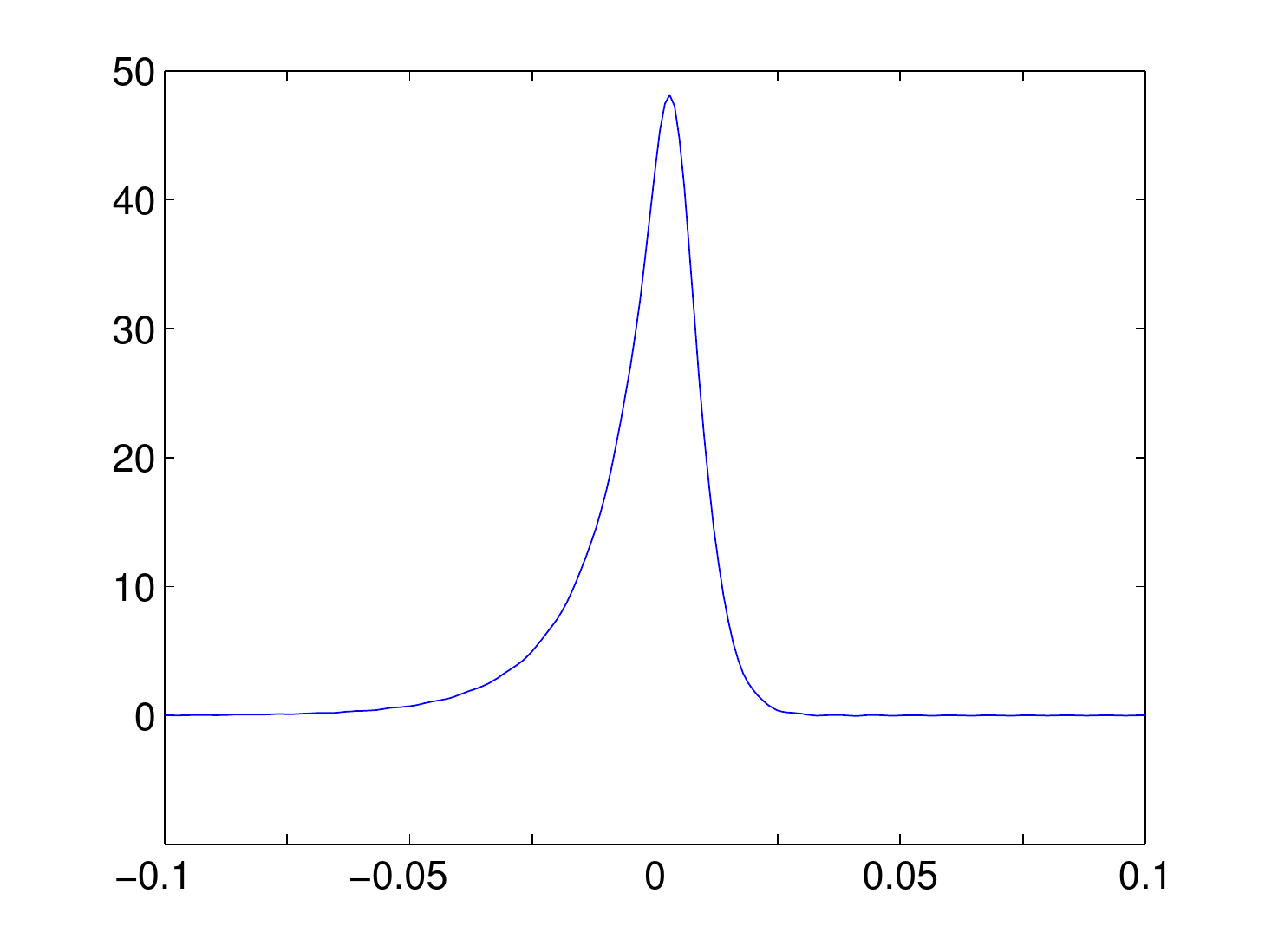}
\caption{Probability density function of the third moment variation}\label{Fig:pdf_tm}
\end{figure}

Figure~\ref{Fig:pdf_tm} shows the probability density functions of the annualized third moment variation derived by PDE~\eqref{Eq:PDE_tm}.
As expected, the distribution of the third moment variation is left skewed.

\begin{table}
\centering
\caption{Numerically computed standardized moments with various hedge number $\beta$ with parameter setting $\mu=0.05$, $\kappa = 18, \theta = 0.1, \gamma = 1, \rho = -0.62$ and $T = 0.1$ years}\label{Table:statistics}
\begin{tabular}{ccccc}
\hline
$\beta$ & mean & std.dev. & skewness & kurtosis \\
\hline
0 & 0.0041 & 0.0964 & -0.4281 & 3.3741  \\
10 & 0.0071 & 0.0861 & -0.3097 & 3.2266 \\
20 & 0.0100 & 0.0766 & -0.1875 & 3.1574 \\ 
30 & 0.0129 & 0.0683 & -0.0671 & 3.1563 \\
40 & 0.0158 & 0.0618 & 0.0600 & 3.1586 \\
50 & 0.0188 & 0.0575 & 0.2296 & 3.2184 \\
60 & 0.0216 & 0.0560 & 0.4275 & 3.2131\\
\hline
\end{tabular}
\end{table}

The local truncation error of the Peaceman and Rachford two-dimensional ADI scheme is known as $\mathcal{O}( (\Delta t)^2 + (\Delta r)^2 + (\Delta v)^2)$.
In other words, the error caused by one iteration of the numerical procedure is bounded by $C((\Delta t)^2 + (\Delta r)^2 + (\Delta v)^2)$ for some constant $C$.
Since the closed form formula for the probability distribution of the hedged portfolio does not exist, 
it is difficult to demonstrate the global truncation error for the entire iteration by comparing the exact solution and the numerical solution.
On the other hand, for the underlying asset alone,
the numerically computed p.d.f. of the return under the present procedure can be compared with the p.d.f. retrieved from the known characteristic function of the Heston-type model:
\begin{equation}
c(\psi) = \frac{ \exp\left\{ -\frac{(\I \psi + \psi^2)\theta}{\xi \coth\left(\frac{\xi T}{2}\right) + \kappa - \I \gamma \rho \psi}  + \frac{\kappa\theta T(\kappa - \I \gamma \rho \phi)}{\gamma^2} + \I \psi \mu T \right\} }
{\left(\cosh \frac{\xi T}{2} + \frac{\kappa - \I \gamma \rho \psi}{\xi}\sinh\frac{\xi T}{2}\right)^{\frac{2\kappa\theta}{\gamma^2}} }\label{Eq:Heston_cf}
\end{equation}
where $\xi= \sqrt{\gamma^2 (\psi^2 + \I \psi) + (\kappa - \I \gamma \rho \psi)^2}$.
In the above formula, $V_0 = \theta$ for simplicity as in the numerical procedure.

\begin{figure}
\centering
\includegraphics[width=0.8\textwidth]{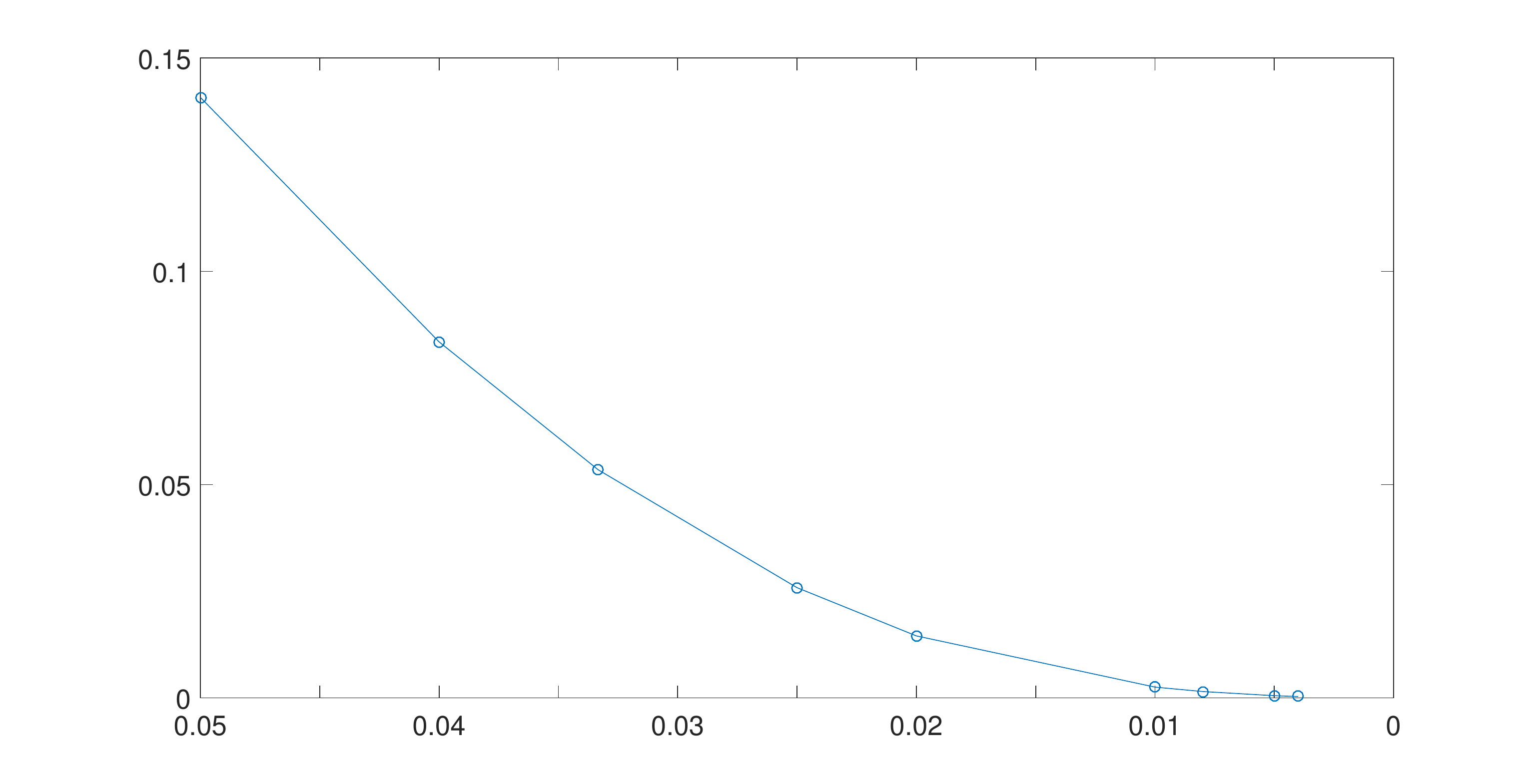}
\caption{Global truncation error with respect to partition size $\Delta r$}\label{Fig:error}
\end{figure}

The result is presented in Figure~\ref{Fig:error}, the global errors measured by the root mean squared errors (RMSE) 
between the numerical PDE solutions and the p.d.f. retrieved from Eq.~\eqref{Eq:Heston_cf} are examined.
For the analysis, the spatial grids are set over $r = [-0.8, 0.8]$ and $v = [0, 0.8]$.
The time step is 0.00001 and the maturity is $T=0.1$.
The global error, which is the total error from whole procedure, is calculated for $\Delta r = 0.05, 0,04, \cdots , 0.005, 0.004$.
The step for variance, $\Delta v$, also changes accordingly such that the number of partitions in $r$ is equal to the number of partitions in $v$.
The result shows that as the partition size decrease, the RMSE converges to zero.
For example, when $\Delta r = 0.004$, the RMSE is $3.19\times10^{-4}$, which is very close to zero.
Since the numerical analysis on the p.d.f. of the hedged portfolio is based on the same method, the numerical solution for the hedged portfolio's return distribution has the same level of accuracy.

\section{Jump diffusion stochastic volatility}\label{Sect:JD}
In this section, we consider a stochastic volatility jump diffusion model with jump in return process:
\begin{align*}
R_t &= \int_0^t \left( \mu - \frac{1}{2}V_s \right) \D s + \int_0^t \sqrt{V_s} \D W^s_s + \sum_{0<s\leq t}\Delta R_s\\
V_t &= \int_0^t \kappa (\theta - V_s) \D s + \int_0^t \gamma \sqrt{V_s} \left(\rho \D W^s_s + \sqrt{1-\rho^2} \D W^v_s\right).
\end{align*}
Then the third moment variation is
$$ [R,R^2]_t = 2 \int_0^t R_{s}V_s \D s + \sum_{0<s\leq t} \Delta R_s \Delta \left( R^2_s \right)$$
and the hedged portfolio return is approximated by
\begin{align*}
X_T &\approx R_T - 2\beta\int_0^T R_{s} V_s \D s - \beta\sum_{0<s\leq t} \Delta R_s \Delta (R^2_s) \\
&= \int_0^t \left( \mu - \frac{1}{2}V_s - 2\beta R_{s} V_s \right) \D s + \int_0^t \sqrt{V_s} \D W^s_s + \sum_{0<s\leq t}\left\{ \Delta R_s - \beta\Delta R_s \Delta \left( R^2_s \right)\right\}\\
&= \int_0^t \left( \mu - \frac{1}{2}V_s - 2\beta R_{s} V_s \right) \D s + \int_0^t \sqrt{V_s} \D W^s_s + \sum_{0<s\leq t}\left\{\Delta R_s - 2\beta R_{s-}\left( \Delta R_s \right)^2 - \beta \left( \Delta R_s \right)^3\right\}
\end{align*}
where we use
\begin{align*}
\Delta R_s - \beta\Delta R_s \Delta \left( R^2_s \right) &= \Delta R_s - \beta \Delta R_s (R_s^2 - R_{s-}^2)\\
&= \Delta R_s - \beta \Delta R_s (R_s - R_{s-})(R_s + R_{s-})\\
&= \Delta R_s - \beta \Delta R_s (R_s - R_{s-})(2R_{s-} + R_s - R_{s-})\\
&= \Delta R_s - 2\beta R_{s-}\left( \Delta R_s \right)^2 - \beta \left( \Delta R_s \right)^3.
\end{align*}

Note that by taking the long-run expectation of the variance $\E [V_t] \approx \theta$, we have
$$\E[ R_{s-}] = \left(\mu -\frac{1}{2}\theta \right)s.$$
By assuming that the jump size $\Delta R_s$ is independent from the current level of return $R_{s-}$, the expectation of the jump part of the portfolio return is represented by an integration form with jump measure $J$.
That is
\begin{align*}
&\E\left[\sum_{0<s\leq t}\left\{\Delta R_s - 2\beta R_{s-}\left( \Delta R_s \right)^2 - \beta \left( \Delta R_s \right)^3\right\} \right]\\ 
&= \int_{[0,t]\times \mathbb R} \left\{ z - 2\beta \left(\mu -\frac{1}{2}\theta \right)s z^2 - \beta z^3 \right\} J(\D s \times \D z)\\
& = \lambda \int_0^t \int_\mathbb R \left\{ z - 2\beta \left(\mu -\frac{1}{2}\theta \right)s z^2 - \beta z^3 \right\} \psi(z) \D z \D s.
\end{align*}
For the last equality, we assume that the jump process follows a normal distribution with density function $\psi$ for jump size and a Poisson distribution with intensity parameter $\lambda$ for the number of jumps over unit time period.
The jump size and arrival time are independent from each other.
In addition, the jump part of the return is assumed to be a martingale, i.e., $\E[\Delta R_s] = 0$ and hence $\psi$ has zero mean.
The equation implies that if there is a jump in the asset return process $R$ with size of $z$ at time $s$, then there is a jump with size of $ z_X(s) := z - 2\beta \left(\mu -\frac{1}{2}\theta \right)s z^2 - \beta z^3$ in the portfolio return process $X$.

Consider a twice continuously differentiable function $h(x,r,v,t)$ such that, for $0 \leq t \leq T$, 
$$ h(x,r,v,t) = \E[H(X_T,R_T,V_T)| x=X_t,r=R_t,v=V_t ]$$
with terminal condition $H(X_T,R_T,V_T)$.
The conditional expectation of the jump size is represented by
$$ \E[\Delta h(X_t, R_t, V_t, t) | X_{t-}, R_{t-},V_t] = \int_{\mathbb R} h(X_{t-} + z_X(t), R_{t-}+z, V_t, t) \psi(z) \D z - h(X_{t-}, R_{t-}, V_t, t).$$
In this point of view, define a differential-integro operator by 
\begin{align}
\mathcal L h(x,r,v,t) ={}&  \left(\mu - \frac{1}{2}v -2\beta r v \right) \frac{\partial h}{\partial x} + \left(\mu-\frac{1}{2}v\right)\frac{\partial h}{\partial r} + \kappa(\theta-v)\frac{\partial h}{\partial v} \nonumber \\
& + \frac{1}{2}v \frac{\partial^2 h}{\partial x^2} + \frac{1}{2}v \frac{\partial^2 h}{\partial r^2}
 + \frac{1}{2}\gamma^2 v\frac{\partial^2 h}{\partial v^2} + \rho \gamma v \frac{\partial^2 h}{\partial x \partial v} + \rho \gamma v \frac{\partial^2 h}{\partial r \partial v} + v \frac{\partial^2 h}{\partial x \partial r}\nonumber\\
&+  \lambda \left( \int_{\mathbb R} h(x + z_X(t),r+z, v, t) \psi(z) \D z - h(x,r,v,t) \right) \label{Eq:L}
\end{align}
and its $L_2$-adjoint, in the sense that $\left< \mathcal L u, w \right> = \left< u, \mathcal L^* w \right>$, for all $u$ and $w$ with inner product $<\cdot,\cdot>$ over $L_2(\mathbb R^3)$, by
\begin{align}
\mathcal L^{*}f(x,r,v,t)={}& -\frac{\partial }{\partial x}\left(\mu - \frac{1}{2}v -2\beta r v \right)f - \frac{\partial}{\partial r}\left(\mu-\frac{1}{2}v\right)f - \frac{\partial}{\partial v}\kappa(\theta-v) f  \nonumber\\
& +  \frac{\partial^2}{\partial x^2}\frac{1}{2}v f +  \frac{\partial^2}{\partial r^2}\frac{1}{2}v f
 + \frac{\partial^2}{\partial v^2}\frac{1}{2}\gamma^2 v f +  \frac{\partial^2}{\partial x \partial v}\rho \gamma v f+  \frac{\partial^2}{\partial r \partial v}\rho \gamma v f + \frac{\partial^2}{\partial x \partial r}v f \nonumber\\
&+ \lambda \left( \int_{\mathbb R} f(x-z_X(t),r-z,v,t) \psi(z) \D z - f(x,r,v,t) \right).\label{Eq:L2}
\end{align}
It is well known that the parts involving derivatives in Eqs.~\eqref{Eq:L}~and~\eqref{Eq:L2} are adjoint to each other.
For the integration part, we show that
\begin{align*}
&\left<\int_{\mathbb R} h(x+z_X(t), r+z, v, t)\psi(z) \D z, f(x,r,v,t) \right>\\
&= \int_{\mathbb R^2}\int_{\mathbb R} h(x+z_X(t), r+z, v, t)f(x,r,v,t)\psi(z) \D z \D x \D r \D v\\
&= \int_{\mathbb R^2}\int_{\mathbb R} h(x', r', v, t)f(x'-z_X(t),r'-z,v,t)\psi(z) \D z \D x' \D r' \D v\\
&= \left< h(x',r',v,t), \int_{\mathbb R} f(x'-z_X(t),r'-z,v,t)\psi(z) \D z \right>.
\end{align*}
Applying It\'{o}'s formula, we have
\begin{align*}
h(X_t, R_t, V_t,t) ={}& \int_0^t \frac{\partial h(X_{s-}, R_{s-}, V_s ,s)}{\partial s} + \mathcal L h(X_{s-}, R_{s-}, V_s ,s) \D s  \\
&+\int_0^t \left( \sqrt{V_s}\frac{\partial h(X_{s-}, R_{s-}, V_s, s)}{\partial x}  + \sqrt{V_s}\frac{\partial h(X_{s-}, R_{s-}, V_s, s)}{\partial r} \right) \D W^s_t \\
&+\int_0^t  \gamma\sqrt{V_s}\frac{\partial h(X_{s-}, R_{s-}, V_s,s)}{\partial v} \D W^v_s \\
&+ \sum_{0<s\leq t} \Delta h -  \lambda t\E[\Delta h(X_t, R_t, V_t, t) | X_{t-}, R_{t-},V_t]
\end{align*}
The last three lines are martingales and we have a backward partial differential-integro equation by setting the integrand of the first line of the above equation to be zero:
$$ \frac{\partial h}{\partial t} + \mathcal L h = 0.$$
In addition, we have a forward equation for the joint probability density function $f$ with the adjoint operator
\begin{equation}
\frac{\partial f}{\partial t} = \mathcal L^{*} f \label{Eq:PDE_JD}
\end{equation}
with initial condition $f(x,r,v,0) = \delta(x)\delta(r)\delta(v-v_0)$.
For further and rigorous information about the Fokker-Planck or forward equation for jump diffusion model, see~\cite{Pappalardo}, \cite{Andersen}, \cite{Hanson}, \cite{bentata2009}, \cite{cont2010}.

As in the previous section, we use the PDE~\eqref{Eq:PDE_JD} to compute the joint probability density function of $(X_t, R_t, V_t)$.
To reduce the dimension, we apply the Fourier transform with respect to $x$, and the transformed PDE is
\begin{align*}
\frac{\partial \hat f}{\partial t} ={}& \left(-\mu + \frac{1}{2}v + \I \phi v + \rho \gamma\right)\frac{\partial \hat f}{\partial r} + \frac{v}{2}\frac{\partial^2 \hat f}{\partial r^2} + \left\{-\kappa(\theta-v) + \gamma^2 + \I \rho \gamma \phi v \right\}\frac{\partial \hat f}{\partial v} + \frac{\gamma^2}{2}v \frac{\partial^2 \hat f}{\partial v^2} \\
&+ \rho \gamma v \frac{\partial^2 \hat f}{\partial r \partial v} + \left\{ \I\phi\left( -\mu + \frac{1}{2}v + 2\beta rv \right) - \frac{\phi^2}{2}v + \I \rho \gamma \phi + \kappa - \lambda \right\} \hat f \\
&+ \lambda \int_{\mathbb R} \e^{-\I z_X(t) \phi} \hat f(r-z,v,t;\phi) \psi(z) \D z .
\end{align*}
By applying the numerical procedure explained in the previous section, the probability density functions are calculated under the stochastic volatility jump diffusion model.

\begin{figure}
\centering
\includegraphics[width=0.45\textwidth]{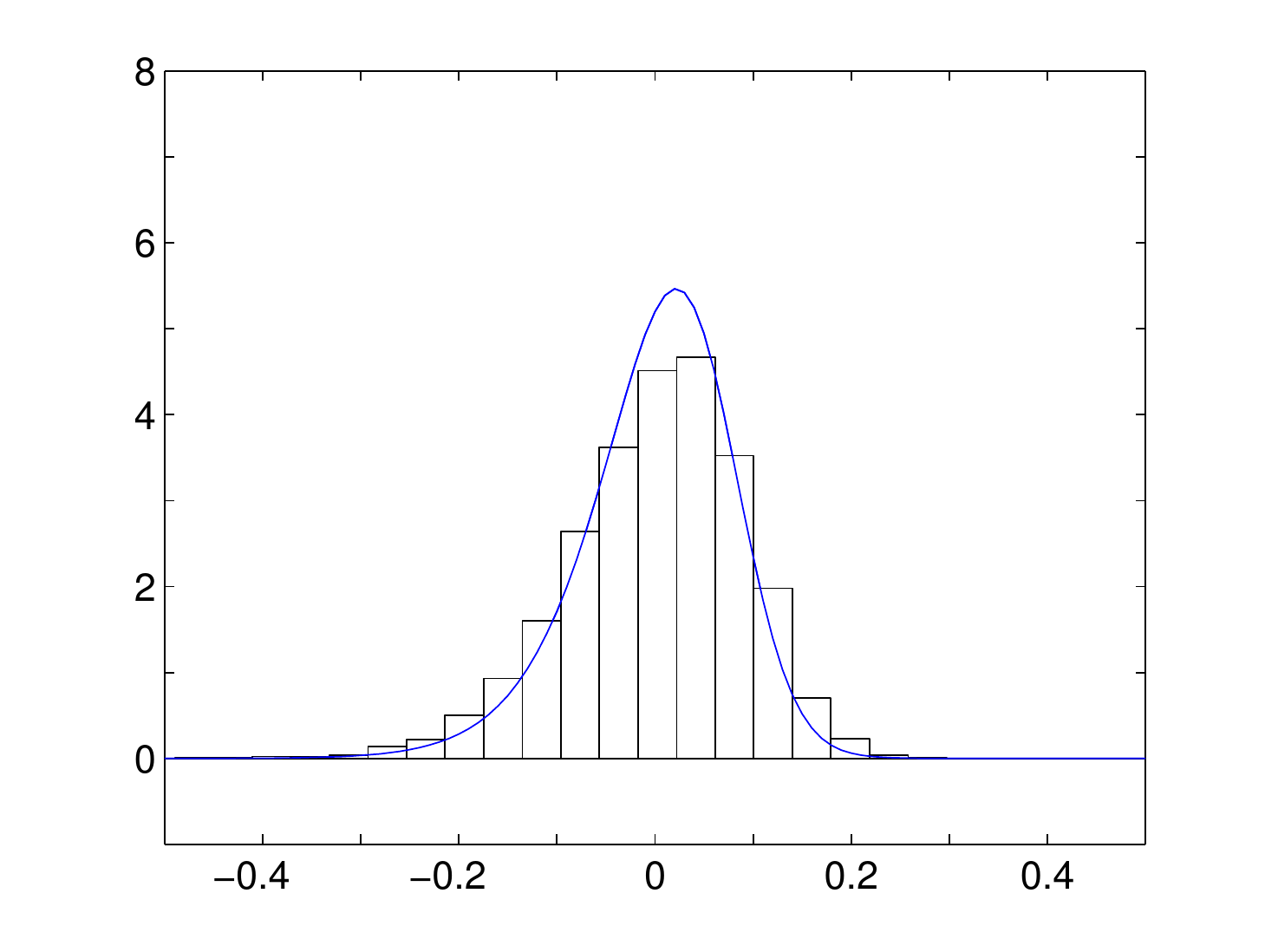}
\includegraphics[width=0.45\textwidth]{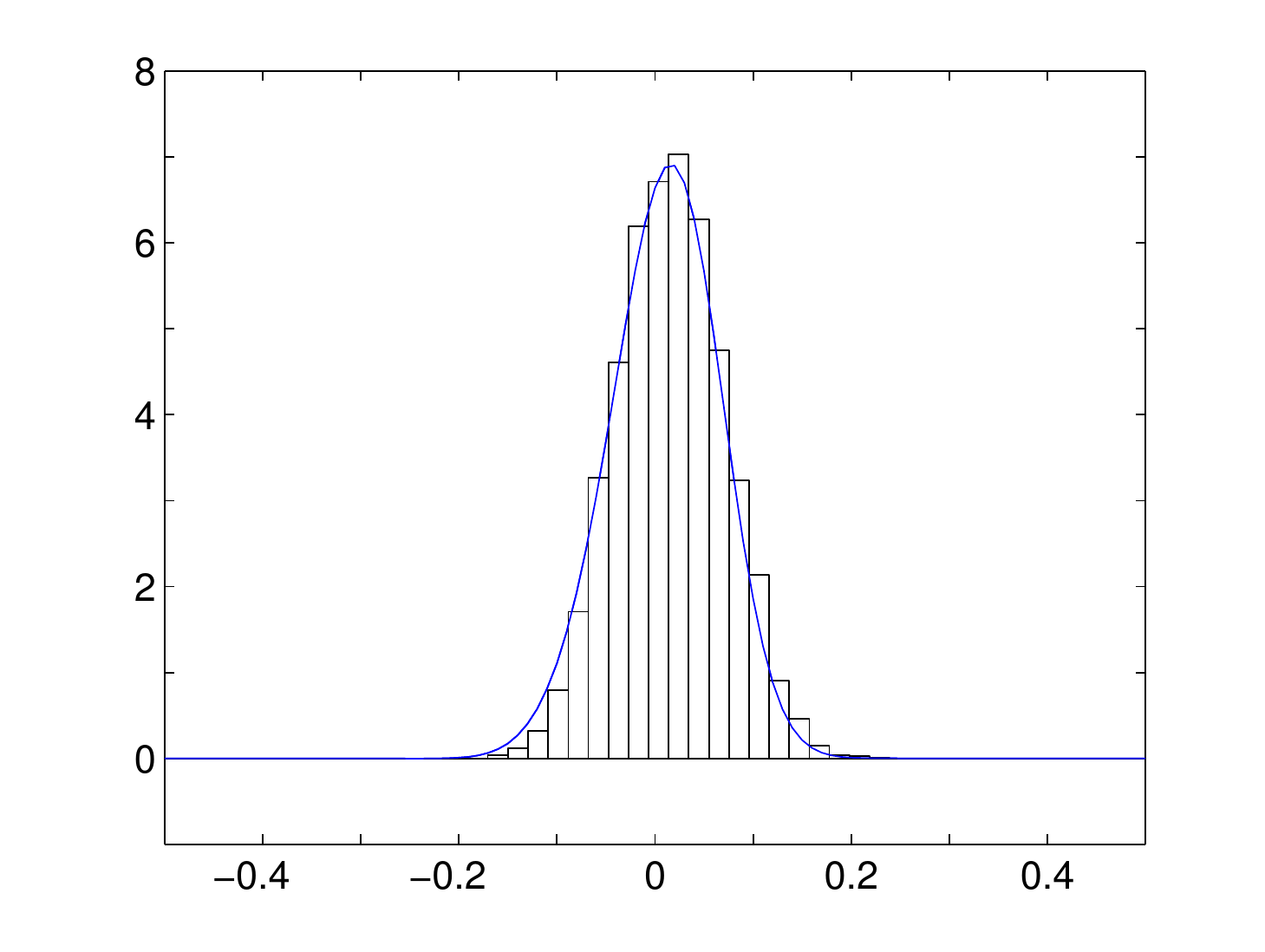}
\caption{Probability density functions and histograms of simulated data under stochastic volatility and jump diffusion : underlying asset (left) and hedged portfolio (right)}\label{Fig:hist_JD}
\end{figure}

\begin{figure}
        \centering
        \begin{subfigure}[b]{0.45\textwidth}
                \includegraphics[width=\textwidth]{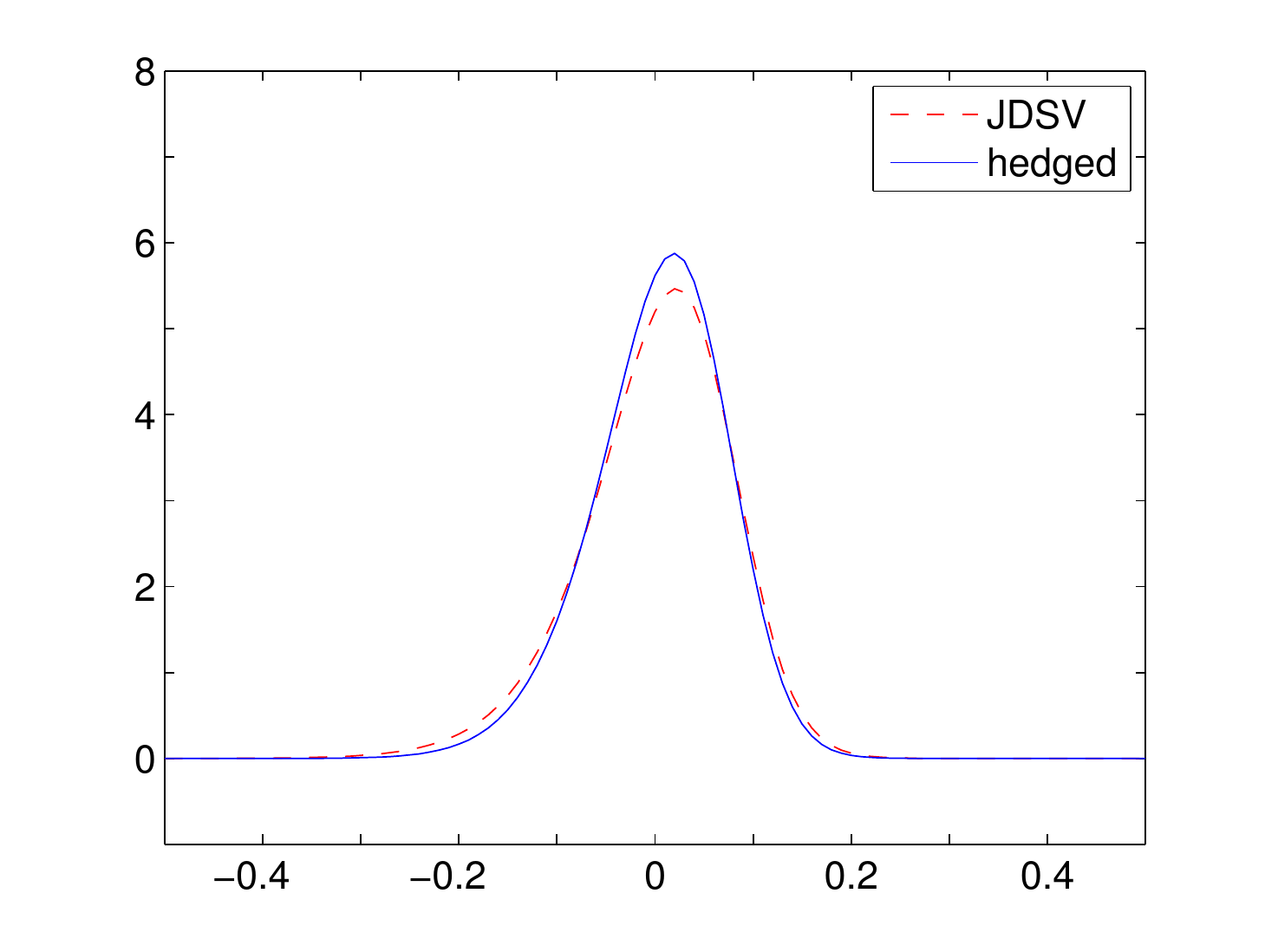}
                \caption{$\beta = 15$}
                \label{Fig:pdf_JD_beta15}
        \end{subfigure}
        \quad  
        \begin{subfigure}[b]{0.45\textwidth}
                \includegraphics[width=\textwidth]{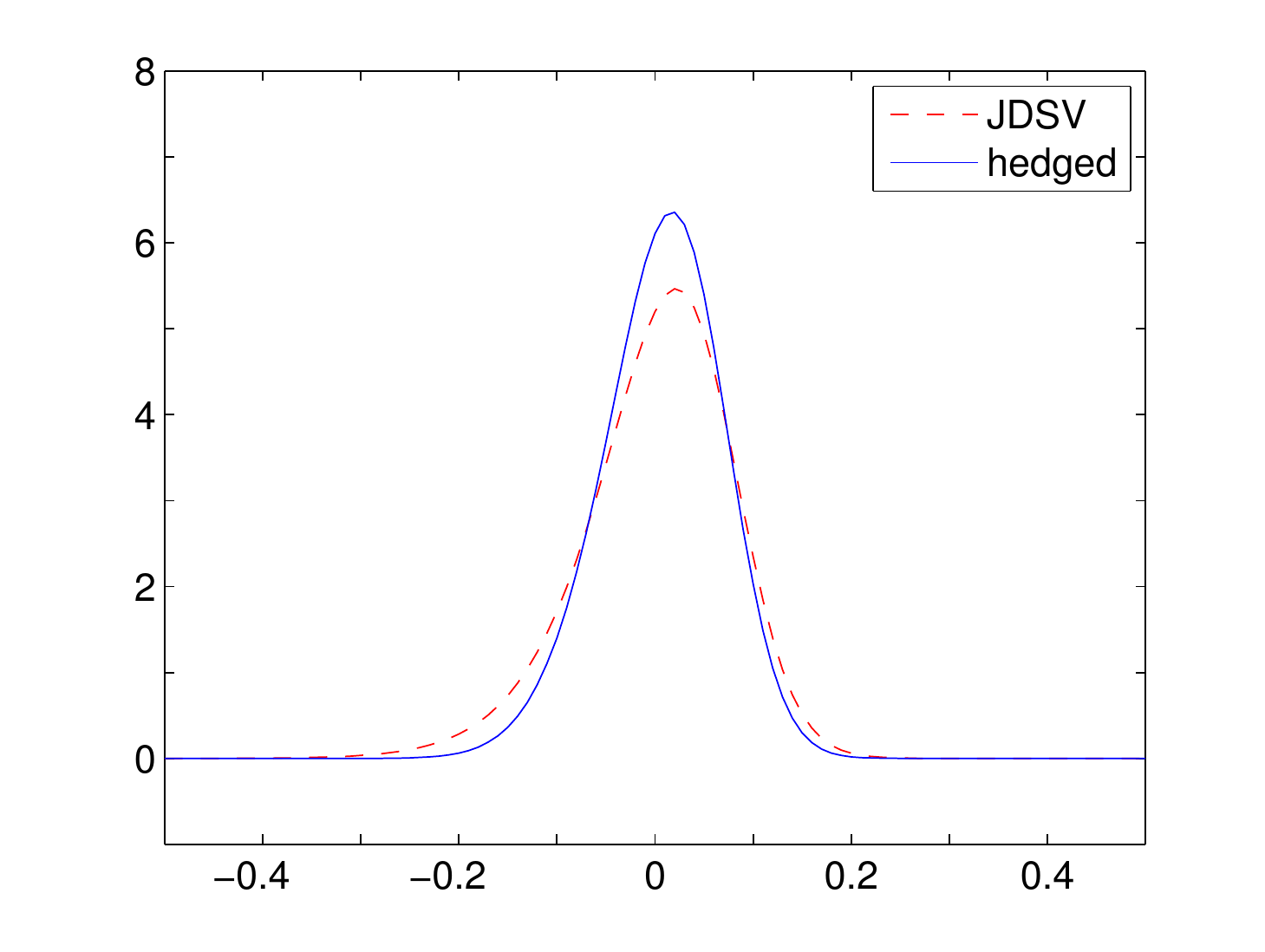}
                \caption{$\beta = 30$}
                \label{Fig:pdf_JD_beta30}
        \end{subfigure}
        
        \begin{subfigure}[b]{0.45\textwidth}
                \includegraphics[width=\textwidth]{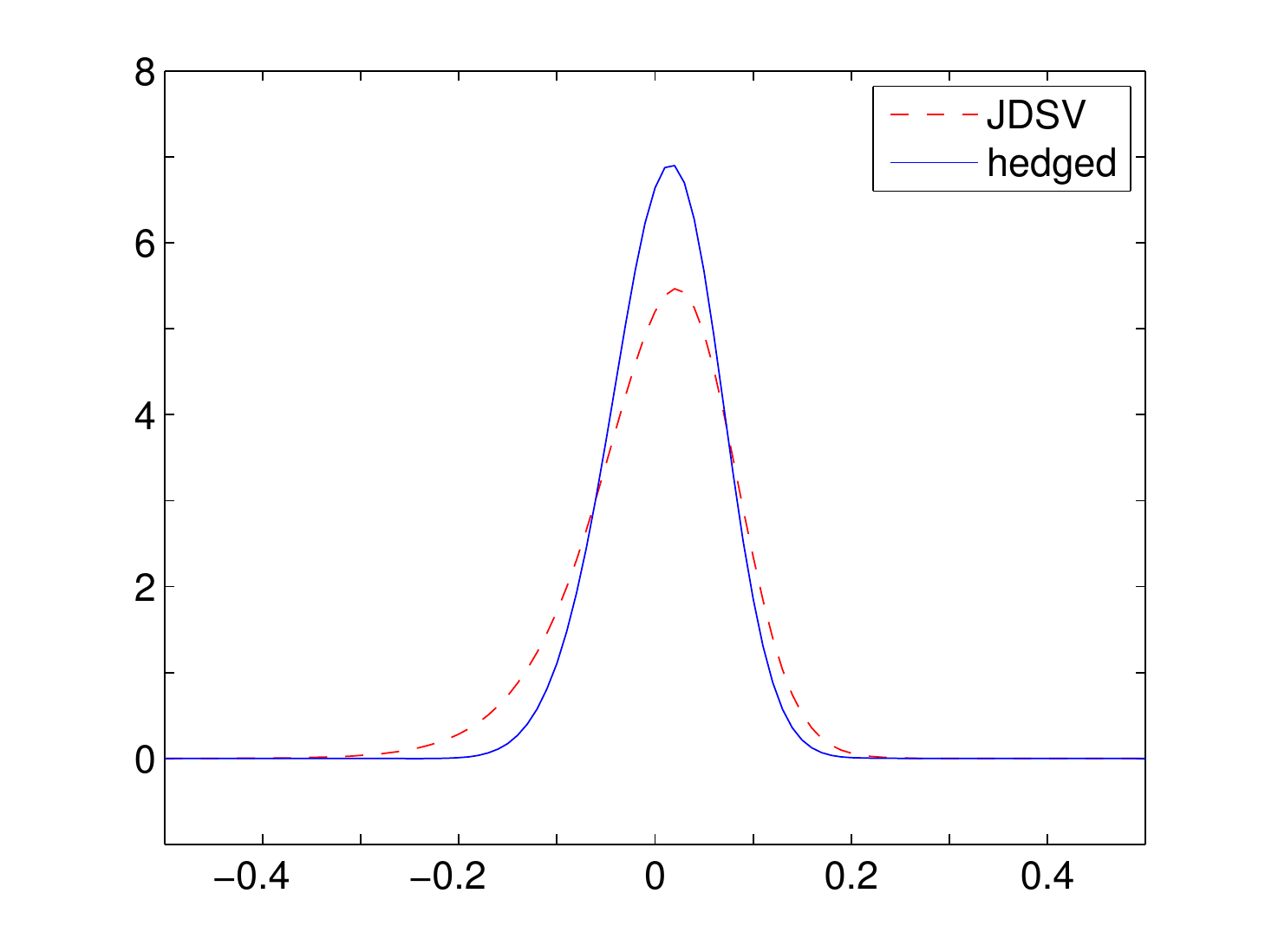}
                \caption{$\beta = 45$}
                \label{Fig:pdf_JD_beta45}
        \end{subfigure}
  		\quad  
        \begin{subfigure}[b]{0.45\textwidth}
                \includegraphics[width=\textwidth]{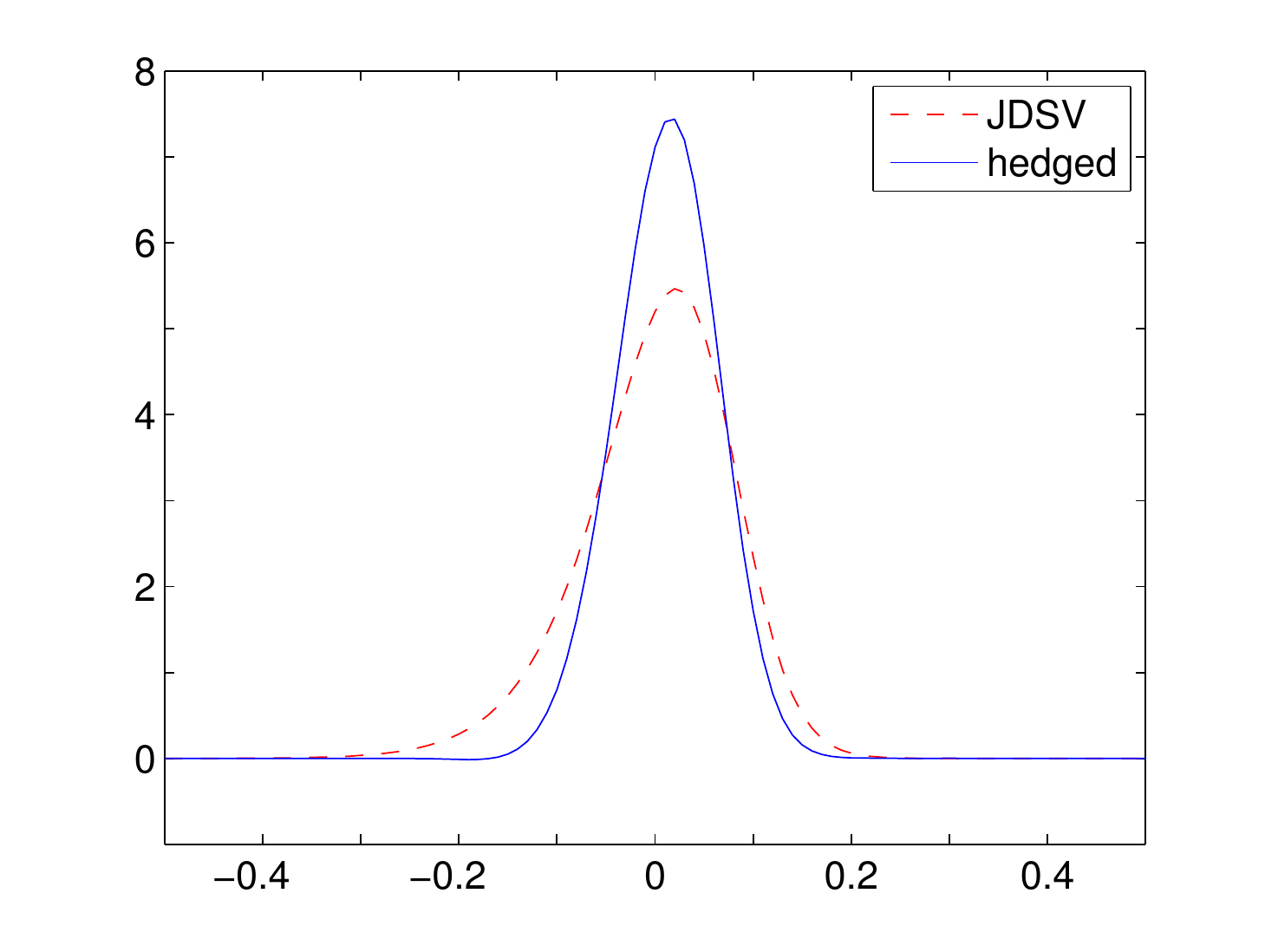}
                \caption{$\beta = 60$}
                \label{Fig:pdf_JD_beta60}
        \end{subfigure}
		\caption{Probability density functions with various hedge numbers $\beta$ (solid) compared to the underlying asset under stochastic volatility jump diffusion (SVJD) model (dashed)}\label{Fig:pdf_JD}
\end{figure}

In this study, $\lambda = 20$ and the standard deviation of jump size $\sigma_j= 0.01$.
For the return and volatility parameters, $\mu=0.05, \kappa = 18, \theta = 0.05, \gamma = 1, \rho = -0.62$.
Figure~\ref{Fig:hist_JD} shows the probability density functions of the underlying asset (left) and the hedged portfolio (right) with the hedge number $\beta = 45$ and $T = 0.1$ compared to the histograms of the simulated data.
The sample size of the simulation is $10^5$.
In Figure~\ref{Fig:pdf_JD} presents the probability density functions of the hedged portfolio returns (solid) compared to the return of the underlying asset (dashed) with various hedge numbers $\beta = 15,30,45$ and $60$.
As in the previous section, the hedged portfolios have more Gaussian-like thin-tail distributions compared to the distribution of the underlying asset.

Table~\ref{Table:statistics} list the numerically computed mean, standard deviation, skewness and kurtosis of the return distributions of the portfolios with various hedge numbers $\beta = 0, 15, 30, 45, 60$.
With $\beta$ between 45 and 60, the skewness of the portfolio return is approximately zero and has minimum kurtosis.
This result is consistent with the simulation study, where the optimal hedge number is reported to be $45.21$.

\begin{table}
\centering
\caption{Numerically computed standardized moments with various hedge number $\beta$ with parameter setting $\mu=0.05$, $\kappa = 18, \theta = 0.05, \gamma = 1, \rho = -0.62$, $\lambda = 20$, $\sigma_j = 0.02$ and $T = 0.1$ years}\label{Table:statistics}
\begin{tabular}{ccccc}
\hline
$\beta$ & mean & std.dev. & skewness & kurtosis \\
\hline
0 & 0.0012 & 0.0759 & -0.5955 & 3.9757  \\
15 & 0.0039 & 0.0682 & -0.4245 & 3.5589 \\
30 & 0.0065 & 0.0612 & -0.2663 & 3.2889 \\
45 & 0.0091 & 0.0552 & -0.1289 & 3.1433 \\
60 & 0.0117 & 0.0506 & 0.0107 & 3.1072 \\
\hline
\end{tabular}
\end{table}

\section{Conclusion}\label{Sect:concl}
The probability density functions of the tail hedge portfolio with the third moment variation swap were calculated.
The method is based on numerical analysis of alternating direction implicit for the partial differential equations of the joint density functions.
The computed density functions show that the swap properly eliminates the skew and fat tail risk of an underlying asset under Heston's stochastic volatility and jump diffusion stochastic volatility models.
In future work, a faster method to calculate the probability density function will be needed because the partial differential equation approach has time complexity.
Therefore, the computed probability function can be used to find the optimal hedge number of the swap to eliminate the skew and tail risks.

\bibliographystyle{apalike}

\end{document}